\documentclass[preprint,trackchanges]{aastex62}

\usepackage{subfigure}
\usepackage{color}
\usepackage{multirow}
\usepackage{amsmath}

\begin{document}

\title{The Hubble Space Telescope's near-UV and optical transmission spectrum of Earth as an exoplanet}

\author{Allison Youngblood}
\affiliation{Laboratory for Atmospheric and Space Physics, 1234 Innovation Dr, Boulder, CO 80303, USA}
\affiliation{NASA Goddard Space Flight Center, Greenbelt, MD 20771, USA}
\email{allison.youngblood@lasp.colorado.edu}

\author{Giada N. Arney}
\affiliation{NASA Goddard Space Flight Center, Greenbelt, MD 20771, USA}
\affiliation{NASA NExSS Virtual Planetary Laboratory, P.O. Box 351580, Seattle, WA 98195, USA}
\affiliation{Sellers Exoplanet Environments Collaboration, NASA Goddard Space Flight Center, Greenbelt, MD 20771, USA}

\author{Antonio Garc\'ia Mu\~noz}
\affiliation{Zentrum f{̈\"u}r Astronomie und Astrophysik, Technische Universit{\"a}t Berlin, Hardenbergstrasse 36, D-10623, Berlin, Germany}

\author{John T. Stocke}
\affiliation{Center for Astrophysics and Space Astronomy, Department of Astrophysical and Planetary Sciences, University of Colorado, UCB 389, Boulder, CO 80309, USA}

\author{Kevin France}
\affiliation{Laboratory for Atmospheric and Space Physics, 1234 Innovation Dr, Boulder, CO 80303, USA}
\affiliation{Center for Astrophysics and Space Astronomy, Department of Astrophysical and Planetary Sciences, University of Colorado, UCB 389, Boulder, CO 80309, USA}

\author{Aki Roberge}
\affiliation{NASA Goddard Space Flight Center, Greenbelt, MD 20771, USA}

\begin{abstract}
We observed the 2019 January total lunar eclipse with the Hubble Space Telescope's STIS spectrograph to obtain the first near-UV (1700-3200 \AA) observation of Earth as a transiting exoplanet. The observatories and instruments that will be able to perform transmission spectroscopy of exo-Earths are beginning to be planned, and characterizing the transmission spectrum of Earth is vital to ensuring that key spectral features (e.g., ozone, or O$_3$) are appropriately captured in mission concept studies. O$_3$ is photochemically produced from O$_2$, a product of the dominant metabolism on Earth today, and it will be sought in future observations as critical evidence for life on exoplanets. Ground-based observations of lunar eclipses have provided the Earth's transmission spectrum at optical and near-IR wavelengths, but the strongest O$_3$ signatures are in the near-UV. We describe the observations and methods used to extract a transmission spectrum from Hubble lunar eclipse spectra, and identify spectral features of O$_3$ and Rayleigh scattering in the 3000-5500 \AA~region in Earth's transmission spectrum by comparing to Earth models that include refraction effects in the terrestrial atmosphere during a lunar eclipse. Our near-UV spectra are featureless, a consequence of missing the narrow time span during the eclipse when near-UV sunlight is not completely attenuated through Earth's atmosphere due to extremely strong O$_3$ absorption and when sunlight is transmitted to the lunar surface at altitudes where it passes through the O$_3$ layer rather than above it.

\end{abstract}

\section{Introduction} \label{sec:intro}

As we approach the era of directly characterizing the atmospheres of Earth-sized exoplanets, considerable preparation is underway to determine how we would recognize signatures of habitability or life from many parsecs away using techniques like transit spectroscopy and direct imaging. A major component of these preparations is the evaluation of biosignatures, the remotely observable features generated by planetary biospheres (e.g. \citealt{DesMarais2002,Sagan1993,Schwieterman2018}). On modern Earth, one of the most important biosignatures is oxygen (O$_2$), which is generated by oxygenic photosynthesis, the dominant metabolism on our planet. Possibly, this is the most productive metabolism that can evolve on any planet \citep{Kiang2007a,Kiang2007b}, and because it is fueled by cosmically abundant inputs like water, carbon dioxide, and starlight, its necessary ingredients should be ubiquitous on habitable planets. 

An important photochemical byproduct of O$_2$ is ozone (O$_3$). Ozone formation initiates via the photolysis of O$_2$:
\begin{equation}
{\rm O}_2 + h \nu (\lambda < 240~\rm{nm}) \rightarrow \rm{O} + \rm{O}
\end{equation}
\begin{equation}
\rm{O} + \rm{O}_2 + \rm{M} \rightarrow \rm{O}_3 + \rm{M}
\end{equation}
In Earth's spectrum, O$_3$ produces prominent spectral features, including the Hartley-Huggins band between 2000-3500 \AA, the broad Chappuis band centered near 6000 \AA, and a prominent infrared feature near 9.6 $\mu$m. The Chappuis band's particular impact on Earth's visible light reflectance spectrum may be potentially diagnostic for detecting exoplanets like modern Earth \citep{KrissansenTotton2018}.  The Hartley-Huggins band at UV wavelengths is the strongest of these ozone features, and it can be apparent in a spectrum at significantly lower O$_2$ levels than modern Earth's atmospheric abundance (21$\%$ of the atmosphere). For instance, during the mid-Proterozoic period (2.0--0.7 billion years ago), the abundance of atmospheric O$_2$ may have only been 0.1$\%$ of the modern atmospheric level \citep{Planavsky2014}, precluding directly detectable oxygen spectral features. However, even at such low oxygen levels, the strong Hartley-Huggins band can still be prominent, allowing remote observers to infer the presence of photosynthetic life on such a planet when that feature is considered in the context of the rest of the spectrum \citep{Reinhard2017}. 

Ozone might even serve as a ``temporal biosignature'' on Earth-like exoplanets. \cite{Olson2018} show that periodic variability in O$_3$ produced by season-driven variability in O$_2$ might produce modulations of ozone's strong Hartley-Huggins absorption feature that could be observed remotely. These modulations are most detectable when oxygen abundances are at lower, mid-Proterozoic-like abundances, because at modern Earth-like abundances, the Hartley-Huggins O$_3$ feature is saturated.  This type of seasonal biosignature might be seen on an exoplanet whose observed system orientation allows seasonal behavior driven by axial tilt and/or orbital eccentricity to be observed. 

In addition to its role as a key biosignature, O$_3$ is important for surface habitability on Earth-like exoplanets because it (and O$_2$) generates a powerful UV shield that protects life on Earth's surface from radiation damage. Prior to the rise of oxygen on Earth, fatal levels of UV radiation at Earth's surface meant life would have had to take refuge under various physical and chemical UV screens \citep[e.g. in the water column, beneath sediments, using robust sunscreen pigments;][]{Cockell1998}. The rise of oxygen in Earth's atmosphere at the start of the Proterozoic geological eon at 2.5 billion years ago marked the ``great oxygenation event''. This was significant both because it signified a major, irreversible redox transition for Earth's atmosphere via the rise of oxygen and ozone, and also because it meant the rise of a robust UV shield that enabled a diversity of organisms to emerge from the water and eventually spread to cover virtually every land surface of the planet. 

The UV shield afforded by ozone has implications for biosignature detection that extend beyond the atmosphere.  Biosignature searches on planets with robust UV protection can target surface reflectance biosignatures such as, e.g., photosynthetic pigments from organisms that produce unusual reflectance spectra. For example, leaf structure generates a step increase at 0.7 $\mu$m called the ``red edge''. This produces a $<$ 10$\%$ modulation in Earth’s disk integrated brightness at quadrature \citep{Montanes2006}. Additionally, there are other biological pigments on Earth that produce their own strong and distinctive reflectivity signatures \citep{Schwieterman2015, Hegde2015}. Unusual reflectance signatures on exoplanets that do not match known abiotic compounds might therefore suggest biological pigments.

Of course, any possible biosignature (including ozone) detected on an exoplanet must be carefully evaluated in the context of the whole planetary environment, and abiotic ``false positive" processes that can generate biosignature mimics without life must be ruled out (e.g. \citealt{Meadows2017}). One way to strengthen the interpretation of true biosignatures is the detection of multiple lines of converging evidence that point to life (e.g. detection of oxygen, ozone, and surface reflectance biosignatures that suggest photosynthesis coupled to non-detection of spectral features that would imply the oxygen or ozone is produced through abiotic photochemical processes). 

Ozone's importance as a biosignature and a habitability-modifier means its very strong UV feature will be an important target for future  observatories capable of sensing UV wavelengths on exoplanets such as, possibly, the LUVOIR\footnote{https://asd.gsfc.nasa.gov/luvoir/} and HabEx\footnote{https://www.jpl.nasa.gov/habex/} future observatory concepts. Thus, it is useful to study this ozone feature on our own planet as an archetype of similar worlds these missions will someday seek elsewhere in the galaxy. 

Lunar eclipses offer an opportunity for Earth-bound observers to observe Earth as if it were a transiting exoplanet, potentially providing a ground truth comparison for models of Earth as an exoplanet. In a lunar eclipse, the moon serves as a mirror, reflecting sunlight that has been filtered through Earth's atmosphere in a transit-like geometry (Figure~\ref{fig:geom}). Similarly, observations of Earthshine, the diffuse sunlight reflected from the Earth onto the unilluminated lunar surface, probe Earth as a directly imaged exoplanet \citep{Palle2010,Robinson2011,GonzalezMerino2013}. 

Many lunar eclipse observations aimed at studying Earth as an exoplanet have been conducted at visible and infrared wavelengths \citep{Palle2009,Vidal-Madjar2010,GarciaMunoz2011,Ugolnikov2013,Arnold2014,Yan2015a,Yan2015b,Kawauchi2018}, revealing the spectral signatures of many gaseous species as well as aerosols. Observations can be obtained in two different phases of an eclipse, the penumbral and umbral phases. During the penumbral phase, the moon passes through the Earth's penumbral shadow (Figure~\ref{fig:geom}), and the illumination of the moon is due partly to direct solar illumination and partly to solar light transmitted through Earth's atmosphere. This phase is most similar to an exoplanet transit observation \citep{Vidal-Madjar2010}. In a partial eclipse, part of the moon passes through the Earth's umbral shadow, and part through the penumbra. In a total or umbral eclipse, all three phases occur over the course of a few hours. Spatially-integrated information about Earth's atmosphere can be gained during a total eclipse, but the lack of direct sunlight is unlike a distant transit observation.  

In this work, we present the first observations of a lunar eclipse with the Hubble Space Telescope (HST) in low-Earth orbit (LEO), and the first near-UV ($\sim$1700-3200 \AA) observations of Earth as a transiting exoplanet. We observed the January 21, 2019 total lunar eclipse, which was also observed by the Large Binocular Telescope's PEPSI instrument at high spectral resolution from $\sim$7500-9000 \AA~\citep{Strassmeier2020}. With HST, we seek features from the Hartley-Huggins band of O$_3$ (2000-3500 \AA), which lies in a spectral region where features from other species are missing and the O$_3$ analysis is therefore more straightforward \citep{Ehrenreich2006}. 

Observing above Earth's atmosphere in LEO greatly simplifies the detection of atmospheric species by removing the presence of additional atmospheric signatures absorbed during the sunlight's path from the top of Earth's atmosphere to a ground-based observatory along the observatory-lunar axis. Past ground-based observations require significant post-processing to remove the direct Earth spectral signatures from the transiting Earth signatures absorbed along the day-night terminator. Although lunar eclipse observations with space observatories do not suffer from this complication, additional complications arise related to HST's capabilities: low spectral resolution in the allowable lunar observing modes\footnote{http://www.stsci.edu/files/live/sites/www/files/home/hst/documentation/\_documents/UIR-2007-01.pdf} and HST's software-driven inability to stably track the apparent motion of an object as close as the moon.

This paper is organized as follows. We describe the observations and data in Section~\ref{sec:obs}, present our Earth transmission spectra in Section~\ref{sec:results}, and compare to Earth model spectra and discuss implications for exoplanet observations in Section~\ref{sec:discussion}. Section~\ref{sec:conclusions} concludes. 

\section{Observations \& Reductions} \label{sec:obs}

\begin{figure}
    \centering
    \includegraphics[width=\textwidth]{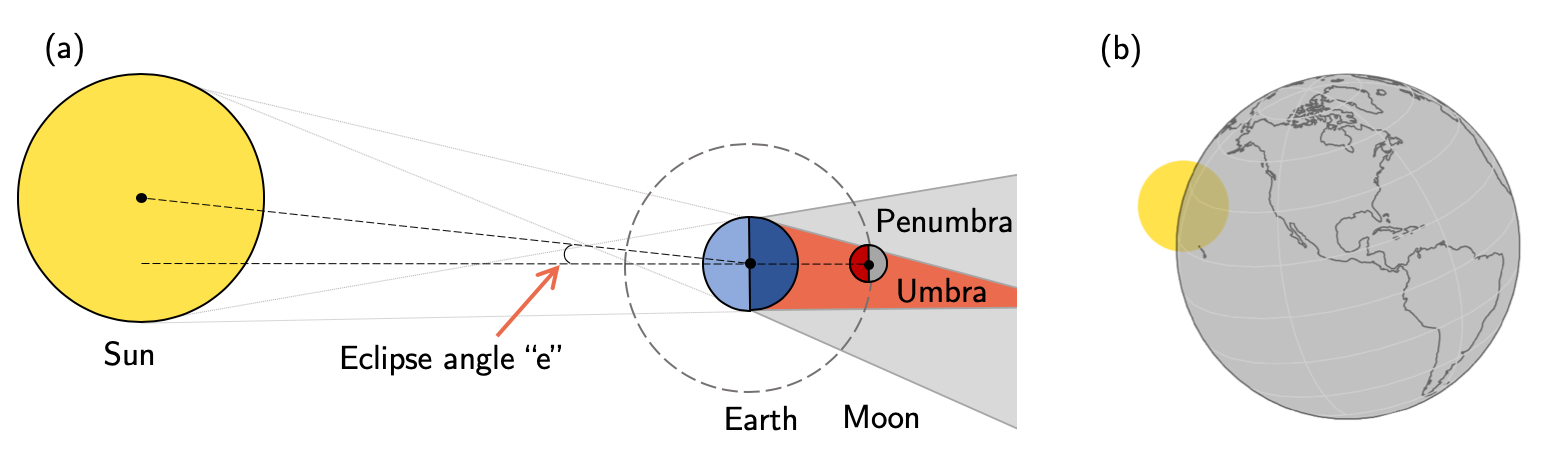}
    \caption{Geometry of the lunar eclipse (all scales exaggerated for visual clarity). \textit{Left panel (a):} Not-to-scale view of the solar system during a lunar eclipse. When the moon is entirely in the Earth's umbra (total or umbral eclipse), all sunlight reaching the lunar surface has been refracted or scattered through Earth's atmosphere. When the moon is in the Earth's penumbra (penumbral eclipse), illumination comes from both unocculted sunlight and sunlight refracted and scattered through the planet atmosphere, similar to an exoplanet transit observation. The Earth-Moon and Earth-Sun axes are drawn and the eclipse angle (or solar elevation angle) $e$ is labelled (see \citealt{GarciaMunoz2011}). \textit{Right panel (b):} A to-scale representation of the Earth and Sun as seen from an observer on the moon at 06:34 UTC on 21 January 2019 during the penumbral eclipse phase. Earth's night time is shaded gray, and the north celestial pole is up and east is left. The Sun's angular distance away from the Earth-moon axis is the eclipse angle $e$. $e\lesssim$0.6$^{\circ}$ corresponds to the umbral phase, and 0.6$^{\circ} \lesssim e \lesssim 1.2^{\circ}$ corresponds to the penumbral phase. }
    \label{fig:geom}
\end{figure}

We observed the January 21, 2019 total lunar eclipse with the Hubble Space Telescope (HST) Space Telescope Imaging Spectrograph (STIS) as part of HST-GO/DD-15674. We were allocated 3 orbits, which we divided between the phases of the eclipse as follows: orbit 1 occurred during the umbral eclipse phase, orbit 2 during the penumbral phase, and orbit 3 out-of-eclipse (see Figure~\ref{fig:geom} and Table~\ref{table:obs}). We utilized 2 different gratings with the STIS CCD and a 52\arcsec$\times$2\arcsec~slit: G230LB (1685-3060 \AA; 1.35 \AA~pix$^{-1}$) and G430L (2900-5700 \AA; 2.73 \AA~pix$^{-1}$). The resolving powers of these modes for a point source are 620-1130 and 530-1040, respectively. To minimize overheads from frequent grating switching, we observed half of each orbit with G230LB and the other half with G430L. We targeted a lunar highlands region (selenographic coordinates: (-8.2$^{\circ}$, -7.3$^{\circ}$)) for its proximity to the center of the moon (to minimize projection effects as well as ensure HST stayed pointed on the lunar surface at all times) and for its proximity to bright highland regions to improve our S/N. A sign error on the submitted HST Phase II form moved our targeted region from the intended (8.2$^{\circ}$, -7.3$^{\circ}$) region to (-8.2$^{\circ}$, -7.3$^{\circ}$). Fortunately, this error still resulted in at least 50\% highlands observations, and HST's pointing remained on the moon during all exposures.

\subsection{Pointing uncertainties} \label{subsec:pointing}
Our observations relied on gyro guiding, because the moon's angular size is larger than HST's field of view and we therefore could not make use of HST's Fine Guidance Sensors (FGS). In order to minimize the initial uncertainty in the pointing on the moon, we must minimize the slew distance between the moon and the last pointing where FGS was used. The closest separation from the moon at which FGS can operate is 9$^{\circ}$, so in the orbit prior to our program's first orbit (pre-orbit 1), a star $\sim$13$^{\circ}$ from the lunar limb was acquired and observations were executed as part of a different program. We assume the initial pointing uncertainty is 14\arcsec. Orbits 1 and 2 of this program were executed sequentially (Visit 02), and after orbit 2 was a mandatory gyro calibration orbit. During this calibration orbit (pre-orbit 3), the FGS acquired a star $\sim$9$^{\circ}$ from the moon, slewed back to the moon, and Orbit 3 was executed (Visit 04).

\begin{figure}
    \centering
    \includegraphics[width=\textwidth]{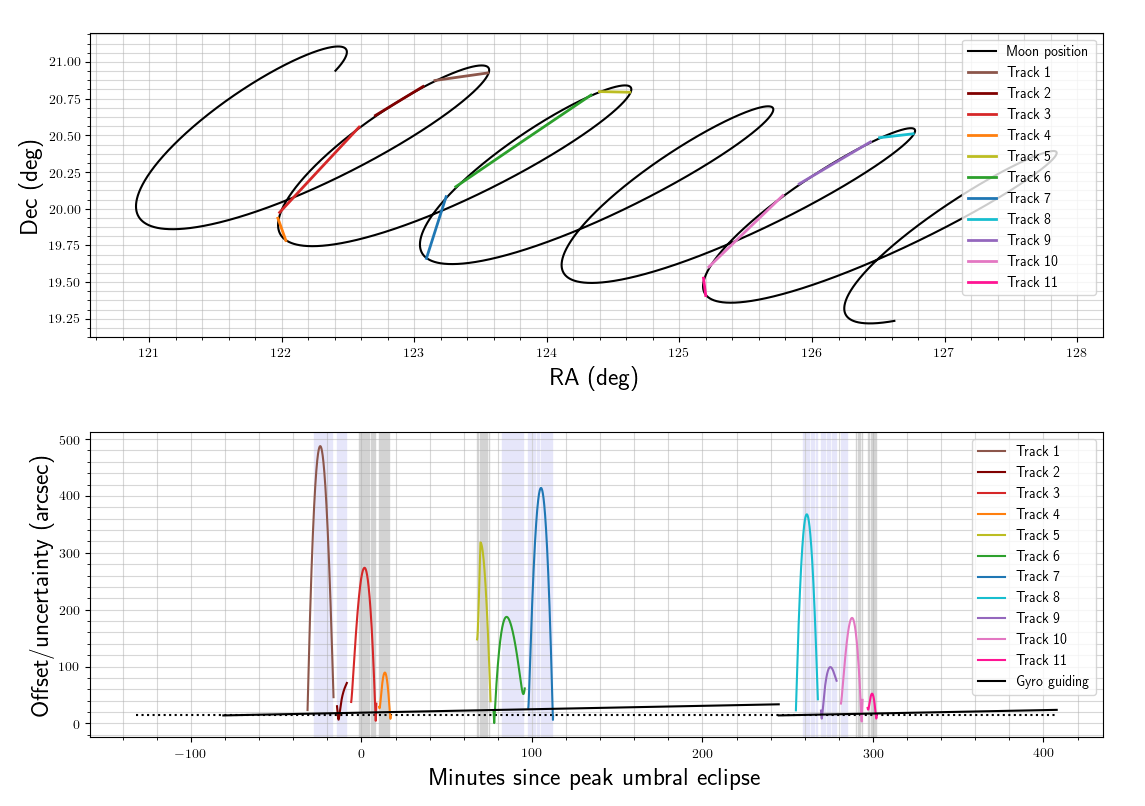}
    \caption{\textit{Top:} The celestial coordinates of our targeted coordinates on the lunar surface (black line) over the course of $\sim$6 HST orbits (95 minutes per orbit) on 21 January 2019. The colored lines show HST's pointing due to the individual linear tracks that were executed (3-4 per orbit). \textit{Bottom:} The pointing offset (due to the linear tracks - colored lines) or pointing uncertainty (due to initial gyro slews - dashed black line -  and gyro guiding - solid black line) is shown against minutes from the peak umbral eclipse time (05:12 UTC). The shaded regions show the time boundaries of each individual exposure. Light blue/purple is for G230LB and light gray is for G430L.}
    \label{fig:pointing_nonlinear}
\end{figure}

\begin{figure}
    \centering
    \includegraphics[width=\textwidth]{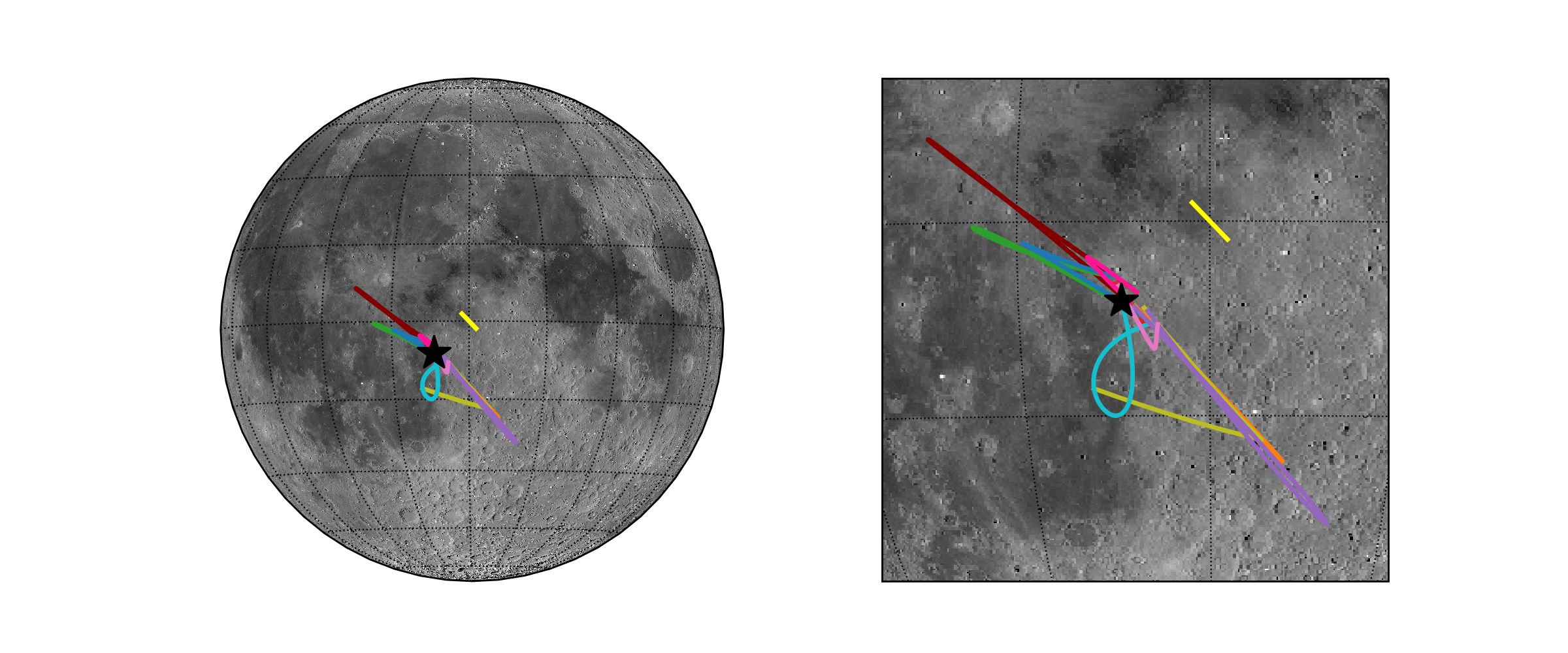}
    \caption{A Lunar Reconnaissance Orbiter Camera (LROC) Wide Angle Camera (WAC) montage of images of the moon taken in the 643 nm bandpass overlaid with the HST pointing positions. The various colors correspond to the color scheme in Figure~\ref{fig:pointing_nonlinear}, and the projected length and approximate orientation of the STIS slit at the equator is shown in yellow. The targeted position (-8.2$^{\circ}$, -7.3$^{\circ}$) is shown with a black star.}    \label{fig:pointing_orthographic}
\end{figure}

Because of the moon's close proximity to Earth and HST in LEO, its apparent motion in the sky is highly non-linear (Figure~\ref{fig:pointing_nonlinear}). HST's software only allows linear tracking of celestial objects, so we had to approximate the moon's non-linear apparent motion with multiple linear tracks. Each initiation of a new linear track requires overhead time that decreases total science observation time. We opted for 3-4 linear tracks per orbit to maximize pointing accuracy while minimizizing overhead per track. We refer the reader to HST User Information Report UIR-2007-001\footnote{http://www.stsci.edu/files/live/sites/www/files/home/hst/documentation/\_documents/UIR-2007-01.pdf} for more details on observing strategies and challenges when observing the moon with HST.

Our program was executed entirely with spectroscopy, so we have no direct information (i.e. from images) of our pointing on the moon. To understand the possible influence of variations in the lunar spectral reflectance on our analysis, we conduct an analysis of our pointing and uncertainty. Our pointing uncertainty comes from three different sources: (1) the initial random pointing error from the gyro slew $\sim$9-13$^{\circ}$ away, (2) gyro guiding uncertainties that accumulate randomly with time, and (3) the absolute error from the linear tracks' approximation of the true motion of the moon (Figure~\ref{fig:pointing_nonlinear}). We assume the initial pointing error is 14\arcsec, and that the gyro guiding uncertainty accumulates at a rate of 0.001\arcsec~s$^{-1}$. In Visit 02, this begins accumulating from the end of pre-orbit 1 to the end of orbit 2, and we estimate the maximum pointing error by the end of orbit 2 is 16.48\arcsec. The true pointing offset will be much less because the gyro guiding error accumulates randomly rather than linearly. The gyro guiding uncertainty resets to zero at the end of pre-orbit 3 and becomes a maximum of 10.72\arcsec~at the end of orbit 3. Finally, the linear tracking approximation goes through a minimum of 0\arcsec~error and a maximum of 10\arcmin~error, dominating the pointing error budget. 

Figure~\ref{fig:pointing_orthographic} shows where HST was pointed on the lunar surface during each of the linear tracks. This pointing was calculated from the celestial coordinates (RA/Dec) of the linear tracks and the NAIF \texttt{SPICE} software \citep{Acton2018}. We mapped the 52\arcsec~STIS slit onto the scale of the lunar disk assuming that the apparent diameter of the lunar disk was a constant 2000\arcsec~over the duration of the observations, which resulted in the scaling 0.09$^{\circ}$/arcsec. In reality, the moon's apparent size ranged from 1968-2039\arcsec~over the course of the observations, translating to a scaling ranging from 0.088$^{\circ}$/arcsec to 0.091$^{\circ}$/arcsec. These differences are likely small compared to uncertainties in the pointing. We determined the orientation of the 52\arcsec~slit for Figure~\ref{fig:pointing_orthographic} using the ORIENT keyword from our data's header files and the angle between the lunar north pole and the north celestial pole from the JPL Horizons Ephemeris System\footnote{https://ssd.jpl.nasa.gov/?horizons}. During Visit 02, the spacecraft roll angle's ORIENT keyword was 283$^{\circ}$ and Visit 04's was 258$^{\circ}$ (we imposed no requirements on ORIENT). The nominal uncertainty on these angles is $\pm$4$^{\circ}$.

\subsection{Reductions} \label{subsec:reductions}

Because of the rapidly changing eclipse conditions between consecutive exposures, we do not stack or combine any of our individual exposures. Beginning with the CalSTIS pipeline's \texttt{flt} data products, which have been subjected to  bias and dark subtraction, flat-fielding, linearity correction, geometric rectification, and wavelength calibration, we applied the L.A.Cosmic \citep{vanDokkum2001} cosmic ray rejection routine. We then ran the \texttt{stistools} x2d function on the cosmic-ray removed \texttt{flt} exposures to get the final 2D spectra. We opt not to absolutely flux calibrate the spectra because of the pipeline's tendency to grossly overcompensate for low-light regions in the G230LB spectra. All transmission spectra with both gratings presented in this work are derived from the ratio of an in-eclipse spectrum (counts s$^{-1}$) to an average of the out-of-eclipse spectra (counts s$^{-1}$), thus cancelling the wavelength-dependent sensitivity function. The upper limits described below are derived from spectra where the absolute flux calibration was applied in the \texttt{stistools} x2d step. 

To extract 1D spectra from our 2D spectra, we median-combined over the CCD's cross-dispersion direction, excluding pixels flagged for the occulting bar, bad or lost data, the overscan region, and saturation. We used a weighted sum of the pipeline errors for the 1-$\sigma$ error bars on the 1D spectra.

In addition to instrumental effects, we need to correct for effects introduced by the varying geometry of sunlight reflection off the Moon. The reflection of sunlight off the moon varies as a function of the incidence, emission, and phase angles and physical properties of the moon's heterogeneous surface \citep{Hapke1981}. Given the large non-linear pointing drifts and uncertainties over the course of individual exposures and HST's rapid motion around the Earth, calculating the requisite angles so that they accurately reflect the mean values for each exposure is challenging. More importantly, many of the terms in the bidirectional reflectance function \citep{Hapke2012} regarding properties of the reflecting medium (the lunar surface) are not well constrained, especially in the near-UV. For simplicity, we assume that reflection off the Moon follows the Lommel-Seeliger function

\begin{equation*}
    F_{LS} = \frac{\cos{\theta_i}}{\cos{\theta_i}+\cos{\theta_e}},
\end{equation*}

\noindent which only depends on the incidence ($\theta_i$) and emission ($\theta_e$) angles, or in other words the angle between the incident sunlight and the lunar surface normal vector and the angle between the telescope boresight direction and the surface normal vector. This equation has been shown to sufficiently approximate the relative reflectivity across the lunar surface \citep{Pettit1930,Hapke1963}. The average incidence and emission angles were calculated from SPICE using the average selenographic coordinates and the mid exposure times reported in Table~\ref{table:obs}, and we find closely coupled incidence and emission angles ranging from 8-30$^{\circ}$ during the eclipse and 6-25$^{\circ}$ out of the eclipse. We also calculated the phase angle, the angle between the telescope boresight direction and the incident sunlight, which ranges from 0.3-1.8$^{\circ}$ during the eclipse and 2.1-3.5$^{\circ}$ out of the eclipse. Using the incidence and emission angles, we find that $F_{LS}$ ranges from 0.4975-0.5010 during the eclipse, and 0.4975-0.5021 out of the eclipse, and each of our 1D spectra were divided by the appropriate $F_{LS}$ value. This small range of $F_{LS}$ values results in $\lesssim$1\% changes to our transmission spectra (calculated as the ratio of the in-eclipse spectra to an average out-of-eclipse spectrum) described in Section~\ref{sec:results}. We also experimented with other viewing angle functions (e.g., $\cos{\theta_i}$ and $\cos{\theta_e}$), and found that the effect was at most $\sim$10\% on our transmission spectra. This level of correction on the transmission spectra would have no impact on the results of this study. 

We estimate the rough magnitude of heterogeneous surface effects on our individual spectra by recording the minimum-maximum range of absolute flux density as a function of wavelength from our out-of-eclipse spectra. In later sections, we include these error bars in our figures in addition to the statistical error bars derived from the pipeline reduction.

Another effect we must correct for is the impact of center-to-limb variations (CLVs) in the solar spectrum on our transmission spectra \citep{Yan2015_CLV}. CLVs arise from the solar atmosphere's temperature-pressure distribution and the fact that light from the solar limb comes from a different atmospheric height than light from disk center. Out of eclipse, the entire solar disk is illuminating the lunar surface, while in eclipse in general only parts of the solar disk are (see right panel of Figure~\ref{fig:geom}), and the Sun exhibits significant, wavelength-dependent CLVs. Therefore, when dividing our in-eclipse spectra by our out-of-eclipse spectra to create transmission spectra, solar features will not completely cancel due to CLVs. This effect is most prominent near the boundary between the penumbral and umbral phases when only the solar limb is illuminating the moon. To correct for CLV effects, we use limb darkening laws from \cite{Eckermann2007} and \cite{Hestroffer1998} for the near-UV and optical, respectively. Ideally, we would use wavelength-dependent limb darkening curves that roughly matched our spectral resolution. However, in the near-UV, large variations have been noted between limb darkening laws \citep{Greve1996} and all available curves introduce considerable noise into our spectra. We elect to use the averaged limb darkening curve of \cite{Eckermann2007}, which is created from an average of various measurements from the literature and is also averaged over wavelengths 2100-3300 \AA, a close match to the G230LB passband. For the optical, we use the limb darkening curve of \cite{Hestroffer1998}, which was also used in \cite{Arnold2014}. Although that work provides spectrally-resolved curves, we use the average curve from the 3030-5500 \AA~range, corresponding to power law index $\alpha$=0.7 and 4110 \AA, to match our approach with the G230LB spectra. For each exposure's mid-time, we calculate the centroid of the visible solar disk above the Earth's limb in order to estimate the average $\mu$ value, which describes the radial position along the solar disk ($\mu$ = $(1-(R/R_{\odot})^2)^{1/2}$ = 1 at disk center and 0 at the limb). For the penumbral phase, these $R/R_{\odot}$ values range 0.7-0.8 for the G430L spectra and 0.001-0.56 for the G230LB spectra. Correspondingly, the limb darkening factors $I/I_0$ range from 0.7-0.79 for the G430L spectra and 0.78-1 for the G230LB spectra. Each penumbral eclipse spectrum is divided by this limb-darkening correction factor in addition to being divided by the Lommel-Seeliger value described above. We do not make this correction for the umbral phase because of the difficulty in calculating the average $R/R_{\odot}$ contributing to illuminating the lunar surface. However, the effect on our transmission spectra should be less than a factor of two, which should have a negligible impact on the conclusions of this study. In later sections we describe residual wavelength-dependent CLVs in our transmission spectra.

In order to determine the effective spectral resolution of our data, we construct a reference solar spectrum from SORCE/SOLSTICE \citep{McClintock2005} and SOLAR-ISS \citep{Meftah2018}. We obtained the SOLSTICE spectrum for Jan 21, 2019 from LISIRD\footnote{http://lasp.colorado.edu/lisird/}, and the SOLAR-ISS spectrum was taken in April 2008 and should be representative of the solar minimum (e.g., conditions akin to Jan 21, 2019). We find good agreement between the SOLAR-ISS spectrum and other spectra (TSIS, OMI, SIM - all from LISIRD) that are lower spectral resolution but were taken on Jan 21, 2019. We convolve the solar spectra with Gaussian kernels of varying widths until it best matches the resolution of our full moon spectra. Based on the width of the best-matching Gaussian kernel, we find that both G230LB and G430L appear to have a spectral resolution of R~$\sim$~100, approximately 6-10$\times$ degraded from the nominal resolution for a point source for these STIS modes.

\begin{figure}
    \centering
    \includegraphics[width=0.5\textwidth]{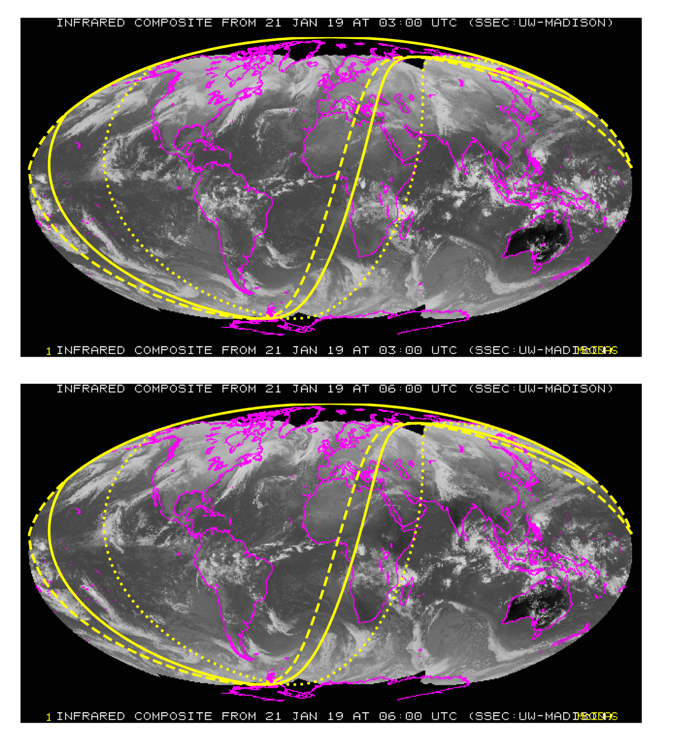}
    \caption{The infrared composite of GOES, Meteosat, and MTSAT data shows cloud cover on 2019-Jan-21 at 03:00 UT (top panel) and 06:00 UT (bottom panel). The yellow day-night terminator lines correspond to 03:00 UT (dotted line), 05:12 UT (solid line), and 06:00 UT (dashed line). Images were downloaded from the University of Wisconsin-Madison's Space Science and Engineering Center\footnote{https://www.ssec.wisc.edu/data/composites/}.}
    \label{fig:weather_map}
\end{figure}

\begin{figure}
    \centering
    \includegraphics[width=0.5\textwidth]{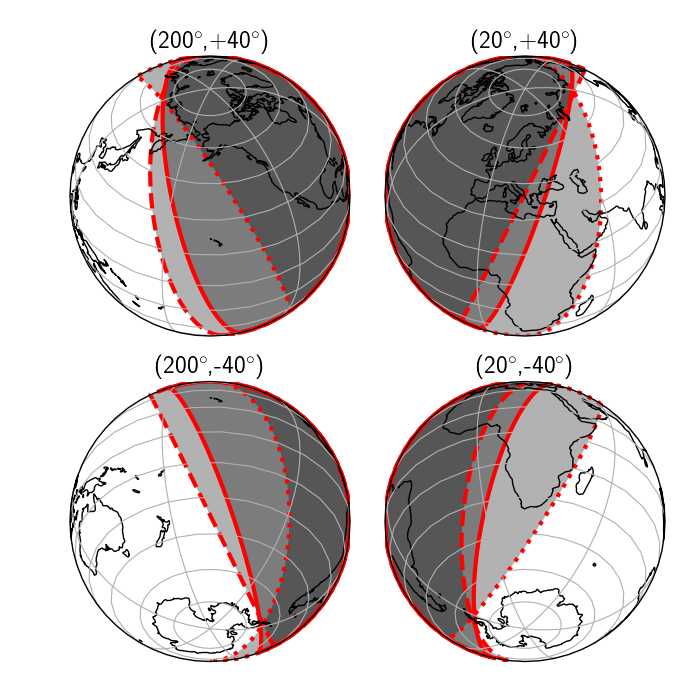}
    \caption{Four views of the Earth with locations of the day-night terminator for 3 different times related to our eclipse observations. The dotted line shows 03:00 UT on 2019-01-21, the solid line shows 05:12 UT (peak umbral eclipse), and the dashed line shows 06:00 UT.  The shaded regions indicates night time.}
    \label{fig:cartopy_4_panels}
\end{figure}

\subsection{Status of space and Earth weather during eclipse} \label{subsec:weather}

We assume that the Sun exhibited no significant time variability over the course of our observations because no X-ray variability and no energetic particles were detected by the GOES satellites. Along the day-night terminator there was significant cloud cover (Figure~\ref{fig:weather_map}). 
{During the penumbral phase, the Sun's rays pass through a very localized area along the terminator, whereas during the deepest parts of the umbral phase ($e$=0.0-0.2$^{\circ}$), rays pass around the entire day-night terminator, with the distribution of rays becoming more concentrated with increasing $e$ (see Figure 2 of \citealt{GarciaMunoz2011}). For the umbral phase of this eclipse, the relevant area of Earth's atmosphere was along the northern latitudes of the terminator shown in Figure~\ref{fig:cartopy_4_panels} (solid line), and during the penumbral phase, the relevant area was over the northwest hemisphere between latitudes $\sim$20-60$^{\circ}$ (i.e., the northern Pacific ocean).

Clouds will affect the overall transmission, especially towards peak umbral eclipse. Aerosols and haze are always present in the atmosphere above the cloud level. They will also affect the overall transmission by increasing the opacity, especially during phases of the eclipse that probe lower altitudes and at short wavelengths where the particle cross sections tend to be largest.

\section{Results} \label{sec:results}

The S/N of our individual eclipse spectra range from $\sim$20-1500 per spectral bin for the penumbral phase with both gratings, and the umbral phase S/N peaked around S/N=60-120 for G430L. The S/N of the out-of-eclipse spectra are generally higher at any given wavelength, except at the bluest end of the G230LB spectra, where S/N $\sim$20. During the umbral eclipse phase, we do not detect photons $\lesssim$4200 \AA, and we present 3$\sigma$~upper surface flux limits (used with the pipeline absolute flux calibration) to aid in planning any future umbral (total) eclipse observations at short wavelengths. To obtain the 3$\sigma$~upper limits, we median-combined each of the flux-calibrated exposures along the cross-dispersion direction as was done for all other spectra described above. We then took the standard deviation of the flux in different wavelength bins and multiplied by three. We then averaged these values across the three G230LB umbral exposures with their standard deviation reported as the uncertainty. We find the surface flux (units erg cm$^{-2}$ s$^{-1}$ \AA$^{-1}$ arcsec$^{-2}$) 3$\sigma$~upper limits to be (1.25$\pm$0.30)$\times$10$^{-14}$ for 1665-2050 \AA, (1.21$\pm$0.05)$\times$10$^{-15}$ for 2050-2550 \AA, and (5.66$\pm$0.08)$\times$10$^{-16}$ for 2550-3070 \AA. Similarly, for the G430L umbral exposures, we find 3$\sigma$~upper limits (5.30$\pm$0.59)$\times$10$^{-16}$ for 2984-3400 \AA, (2.64$\pm$0.42)$\times$10$^{-16}$ for 3400-3800 \AA, and (1.18$\pm$0.10)$\times$10$^{-16}$ for 3800-4200 \AA. Photons at all wavelengths are detected in all of our other spectra. 

We create Earth transmission spectra by dividing our in-eclipse spectra by an average of the full moon spectrum. Both full moon and in-eclipse spectra contain the solar spectrum and lunar albedo signatures, but only the in-eclipse spectra contain the Earth's absorption signature. Thus, dividing the eclipse spectra by the full moon spectra results in a pure Earth spectrum (transmission spectrum), assuming the lunar albedo and solar spectrum signatures completely cancel out.  The low spectral resolution of our data means that the solar features appear to easily cancel out compared to ground-based, high resolution spectroscopy where relative Doppler shifts over the course of the night can have a significant effect (e.g., \citealt{Vidal-Madjar2010}). However, our transmission spectra are equally as affected by solar CLVs as ground-based transmission spectra, and we discuss the impact of residual CLVs on our data in the next section.

Because HST's pointing is less precise and accurate than ground-based telescopes and therefore the STIS slit was not positioned on the same lunar region in the in- and out-of-eclipse spectra, lunar albedo features generally do not cancel as easily as in ground-based spectra. Because over the course of a given exposure, the pointing can change significantly due to long exposure times and large pointing drifts, we generally do not calibrate our penumbral transmission spectra into effective height of the atmosphere as was done in \cite{Vidal-Madjar2010} and \cite{Arnold2014}, except for a comparison with several model spectra at the end of Section \ref{sec:discussion}. In this context, the effective height $h(\lambda)$ of the atmosphere is the altitude under which the atmosphere is considered completely opaque (see Section 4 of \citealt{Vidal-Madjar2010}):

\begin{equation*}
    F_{in} = F_{out} \times \frac{S - L \times h(\lambda)}{S_{\odot}}.
\end{equation*}

\noindent Finding the effective height requires a calculation of the fractional area of the solar surface blocked by the Earth ($S/S_{\odot}$) and the path length of the Earth's limb over which the Sun appears ($L$), and the erratic pointing drift over individual exposures makes that calculation complicated. We perform this calculation for several of our high-S/N penumbral phase optical transmission spectra without obvious evidence of residual solar features, and we fix the effective height to be 23 km between 4520-4540 \AA~as was done in \cite{Arnold2014}. The fractional area of the solar surface blocked by the Earth and path length along Earth's limb intersected by the Sun were calculated by assuming the Sun and Earth are perfect spheres and finding via the JPL Horizons Ephemeris Service the apparent celestial coordinates and apparent sizes of the Sun and Earth from an observer at the center of the moon as a function of time during the eclipse. See Section~\ref{sec:discussion} for a discussion of these calibrated transmission spectra. 

\begin{figure}
    \centering
    \includegraphics[width=\textwidth]{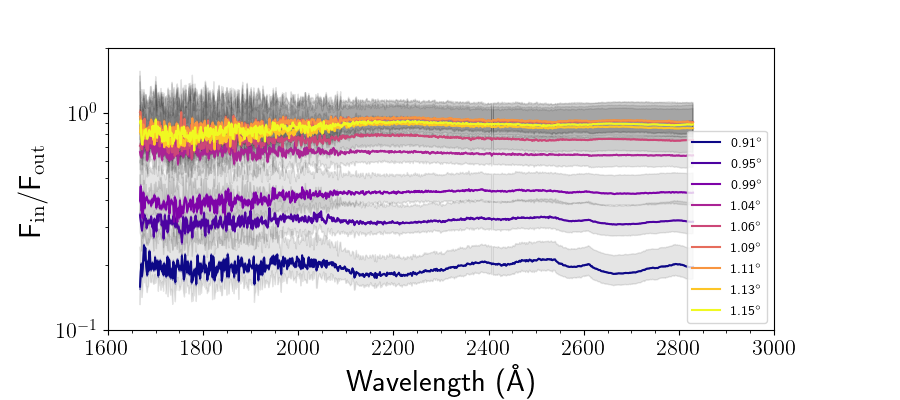}
    \caption{Near-UV transmission spectra (in-eclipse spectra divided by an average of the out-of-eclipse spectra) from the penumbral eclipse phase. The colors correspond to different exposures taken at different eclipse angles with smaller angles corresponding to deeper eclipse times, the colored shaded region represent the statistical uncertainty in the colored lines, and the gray shaded regions represent uncertainty due to varying albedo (based on the range of albedos observed in our out-of-eclipse spectra). }
    \label{fig:G230LB_penumbral_transmission}
\end{figure}

\begin{figure}
    \centering
    \includegraphics[width=\textwidth]{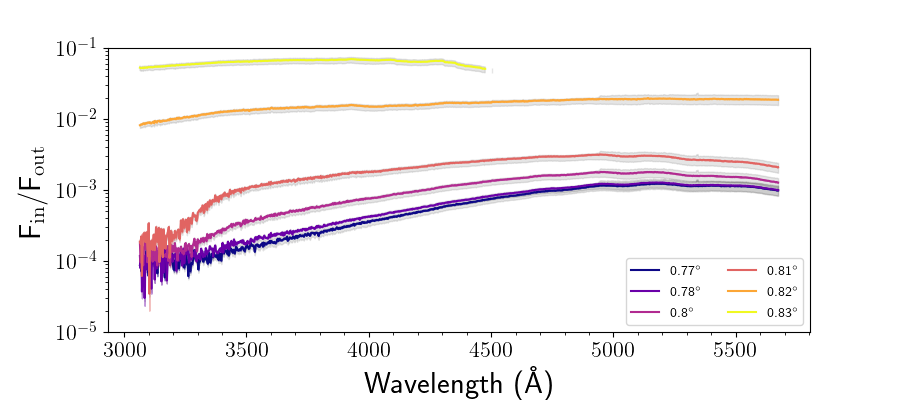}
    \caption{Same as Figure~\ref{fig:G230LB_penumbral_transmission} for the optical spectra of the penumbral eclipse phase. Spectral bins affected by saturation are not shown (i.e., the $e$=0.83$^{\circ}$ spectrum at $>$4500 \AA).}
    \label{fig:G430L_penumbral_transmission}
\end{figure}

\begin{figure}
    \centering
    \includegraphics[width=\textwidth]{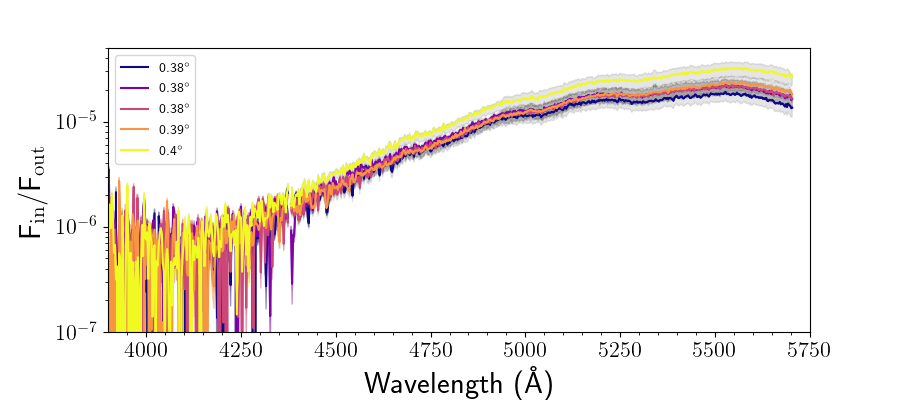}
    \caption{Same as Figure~\ref{fig:G230LB_penumbral_transmission} for the optical spectra of the umbral (total) eclipse phase.}
    \label{fig:G430L_umbral_transmission}
\end{figure}

Figures~\ref{fig:G230LB_penumbral_transmission}-\ref{fig:G430L_umbral_transmission} show our transmission spectra for the penumbral and umbral phases. We obtained 9 spectra with G230LB and 6 spectra with G430L during the penumbral phase, and 3 spectra with G230LB and 5 spectra with G430L during the umbral phase.  Table~\ref{table:obs} lists the details of each exposure, including the mean eclipse angles (the exterior vertex of angle of the Sun-Earth-Moon triangle; Figure~\ref{fig:geom}; \citealt{GarciaMunoz2011}) and the mean longitude and latitudes of the linear tracks. Smaller angles $e$ correspond to times when the moon is deeper in eclipse (deeper in Earth's shadow). 

We detected no photons for $\lambda \lesssim$4200 \AA~during the umbral phase (see upper flux limits above). In the G430L umbral and penumbral spectra, we clearly detect O$_3$ absorption features of the Chappuis band at $>$4000 \AA, as well as the Rayleigh scattering slope at $<$5000 \AA. In the G430L penumbral spectra, we detect broad O$_3$ absorption around 3000-3400 \AA~due to the Hartley-Huggins band. This feature is only apparent at intermediate eclipse angles. The spiky features in some of the penumbral spectra around 3000-3500 \AA~are consistent with noise; the peaks and troughs are only 1-2 spectral bins wide and their wavelengths do not coincide across different transmission spectra. In the G230LB penumbral spectra, we detect three broad features at 2200 \AA, 2550 \AA, and 2675 \AA, at high significance, which appear to be residual solar features (e.g., \ion{Fe}{2}; see Section~\ref{sec:discussion}).

\section{Discussion} \label{sec:discussion}

To interpret our Earth transmission spectra, we compare with the model transmission spectra from \cite{GarciaMunoz2011} that incorporate refraction effects relevant to lunar eclipses into a 1D spherically symmetric model of Earth's atmosphere. To identify spectral features in our data, we compare to a cloud-free and aerosol-free version of the model spectra with only the refracted (``direct") component of sunlight included. Based on \cite{GarciaMunoz2011}, we conservatively assume that contribution of the forward-scattered (``diffuse") component of transmitted light incident on the lunar surface is F$_{\rm in}$/F$_{\rm out}$=10$^{-6}$, roughly independent of wavelength. Only our G430L umbral transmission spectra (Figure~\ref{fig:G430L_umbral_transmission}) approach the 10$^{-6}$ normalized irradiance level at the blue end of the spectra. Our G230LB umbral data are a non-detection, and our upper limits on the normalized irradiance level in the near-UV are $\lesssim$10$^{-5}$. Thus we conclude that the diffuse component, which carries information about the size, shape, and composition of aerosols in the atmosphere, is not significantly affecting our data.

Model transmission spectra are available as a function of the eclipse angle (Figure~\ref{fig:geom}), and the model $e$ values are computed for constant values of the separation between the centers of the Earth and Sun ($a_{\odot}$ = 1 au) and the separation between the center of the Earth and the surface of the Moon ($d_0$ = 382665.9 km). In reality, $a_{\odot}$ and $d_0$ change among various lunar eclipses as well as change during a single eclipse, but these resulting variations in eclipse angle are negligible compared to the variations due to the changing position on the lunar surface during a single observation. We note that the eclipse angles covered by our umbral observations (0.38$^{\circ}$-0.46$^{\circ}$) are probing altitudes $\sim$3-13 km in Earth's atmosphere, and the penumbral observation angles (0.77$^{\circ}$-1.15$^{\circ}$) are probing $\sim$8 km and above \citep{GarciaMunoz2011}. The model does not include clouds or hazes, which should have non-negligible effects at these altitudes, especially during the umbral eclipse. 

To compare our model results and observed spectra, in Figures~\ref{fig:G430L_penumbral_transpec_withmodel}-\ref{fig:G230LB_penumbral_transpec_withmodel}, we overplot the model transmission spectra and the observed transmission spectra. To highlight the spectral contribution of O$_3$, we show two versions of the model spectra: an atmosphere with O$_3$, O$_2$, and Rayleigh scattering, and one without O$_3$. 

As shown in Figure \ref{fig:G430L_penumbral_transpec_withmodel}, our G430L penumbral and umbral transmission spectra show clear evidence of the O$_3$ Chappuis band ($>$4000 \AA). Around e=0.8$^{\circ}$, we detect a drop-off feature around 3300 \AA~attributable to the long-wavelength end of the O$_3$ Hartley-Huggins bands, which impedes UV photons from reaching the ground. O$_3$ and Rayleigh scattering are the dominant features in these spectra. In order to match the model spectra to the observed transmission spectra, we must allow the model eclipse angle to be flexible. This is to account for the uncertainty in our average eclipse angle calculations for each spectrum, and for the opacity-increasing effect of aerosols and clouds in the atmosphere, which are missing from the model, but also for uncertainties in the lunar reflectance, the viewing angle correction, and the CLV correction. For example in Figure~\ref{fig:G430L_penumbral_transpec_withmodel}, the $e$=0.2$^{\circ}$ model spectrum best matches the $e$=0.4$^{\circ}$ data in absolute F$_{\rm in}$/F$_{\rm out}$ values, but it shows significantly steeper slopes toward shorter wavelengths. The $e$=0.3$^{\circ}$ model best matches the spectral features and slope of the $e$=0.38$^{\circ}$ data, but is a factor of 4 larger in F$_{\rm in}$/F$_{\rm out}$. Indeed, aerosols and hazes will affect the eclipse brightness, especially for the smaller eclipse angles as has been shown before \citep{Keen1983, GarciaMunoz2011, GarciaMunoz2011b, Ugolnikov2013}. Aerosols will be vertically stratified, and their absorption signal will depend on wavelength in a way that is difficult to anticipate. Therefore the models will always result in F$_{\rm in}$/F$_{\rm out}$ values larger than the observations. However, it is possible that much of this factor of 4 difference between the observations and model could be caused by CLVs. We did not correct our umbral transmission spectra for CLV effects, because of challenges in estimating the dominant part of the solar disk contributing to illuminating the solar surface during the umbral phase.

The model transmission spectra corresponding to our near-UV penumbral observations are essentially flat and featureless  (Figure~\ref{fig:G230LB_penumbral_transpec_withmodel}). At the largest eclipse angles ($e\sim1.2^{\circ}$), the $F_{in}/F_{out}$ ratio is $\sim$1, indicating that the sunlight is barely attenuated by the Earth's atmosphere at this late stage in the penumbral eclipse. At the shortest wavelengths in Figure~\ref{fig:G230LB_penumbral_transpec_withmodel} are the Schumann-Runge bands of O$_2$ and strong O$_3$ absorption, however our data are noisy in this region and do not show absorption from either species here. We do not detect any of the low-amplitude O$_3$ features or Rayleigh scattering between 2000-2800 \AA, in large part because our transmission spectra are dominated by residual solar spectral features due to unaccounted for wavelength-dependent CLVs. In Figure~\ref{fig:G230LB_residuals_straight_lines}, we compare our near-UV penumbral transmission spectra with a straight line, and we see statistically significant departures from a featureless spectrum at the native data resolution around 2200 \AA~and 2675 \AA. These features are deep in absorption at lower $e$ angles (0.9$^{\circ}$) but fill in toward emission at larger $e$ angles (1.2$^{\circ}$) , which is consistent with how CLV residual features manifest \citep{Yan2015_CLV}. In Figure~\ref{fig:G230LB_residuals_straight_lines}, we overplot the solar spectrum to show that these features appear to have a solar origin, such as Fe$^{+}$, and some higher ionization states of Fe. Besides these solar residuals, we do not detect any spectral features in our near-UV data.

While O$_3$ cross sections are larger in the near-UV than in the optical, there is only a narrow period of time during an eclipse when it is possible to detect near-UV O$_3$ features. For example, there is a brief range of eclipse angles between 0.6-0.8$^{\circ}$ where the excess absorption due solely to O$_3$ is $>$0.1-1\%. This is dictated by the relatively sharp altitude cutoff of O$_3$ in the stratosphere. In the right panel of Figure~\ref{fig:G230LB_penumbral_transpec_withmodel}, the 0.7$^{\circ}$ and 0.8$^{\circ}$ models are shown, while the 0.6$^{\circ}$ is off the scale at the bottom of the plot (i.e., F$_{in}$/F$_{out}$ $\ll$10$^{-6}$). At e$<$0.6$^{\circ}$, no photons reach the lunar surface, because they are all completely absorbed over their refraction path length, a consequence of the extremely strong O$_3$ absorption in the near-UV. The Chappuis bands in the optical are much weaker, and these features are detected at small eclipse angles, even during the umbral eclipse. At e$>$0.8$^{\circ}$, the O$_3$ absorption at all wavelengths becomes very small because the altitudes transmitting sunlight to the lunar surface lie above the O$_3$ layer, and the weak spectral features are difficult to detect without higher precision data. Also, around e=0.6-0.7$^{\circ}$ the normalized irradiance of the models approaches 10$^{-6}$, so potentially the near-UV ozone signature could be swamped by forward-scattered (``diffuse") sunlight off of high-altitude gas or aerosols. However, at near-UV wavelengths, aerosols forward scatter less efficiently than at long wavelengths, so the diffuse component could be $\ll$10$^{-6}$.

\begin{figure}
    \centering
    \includegraphics[width=\textwidth]{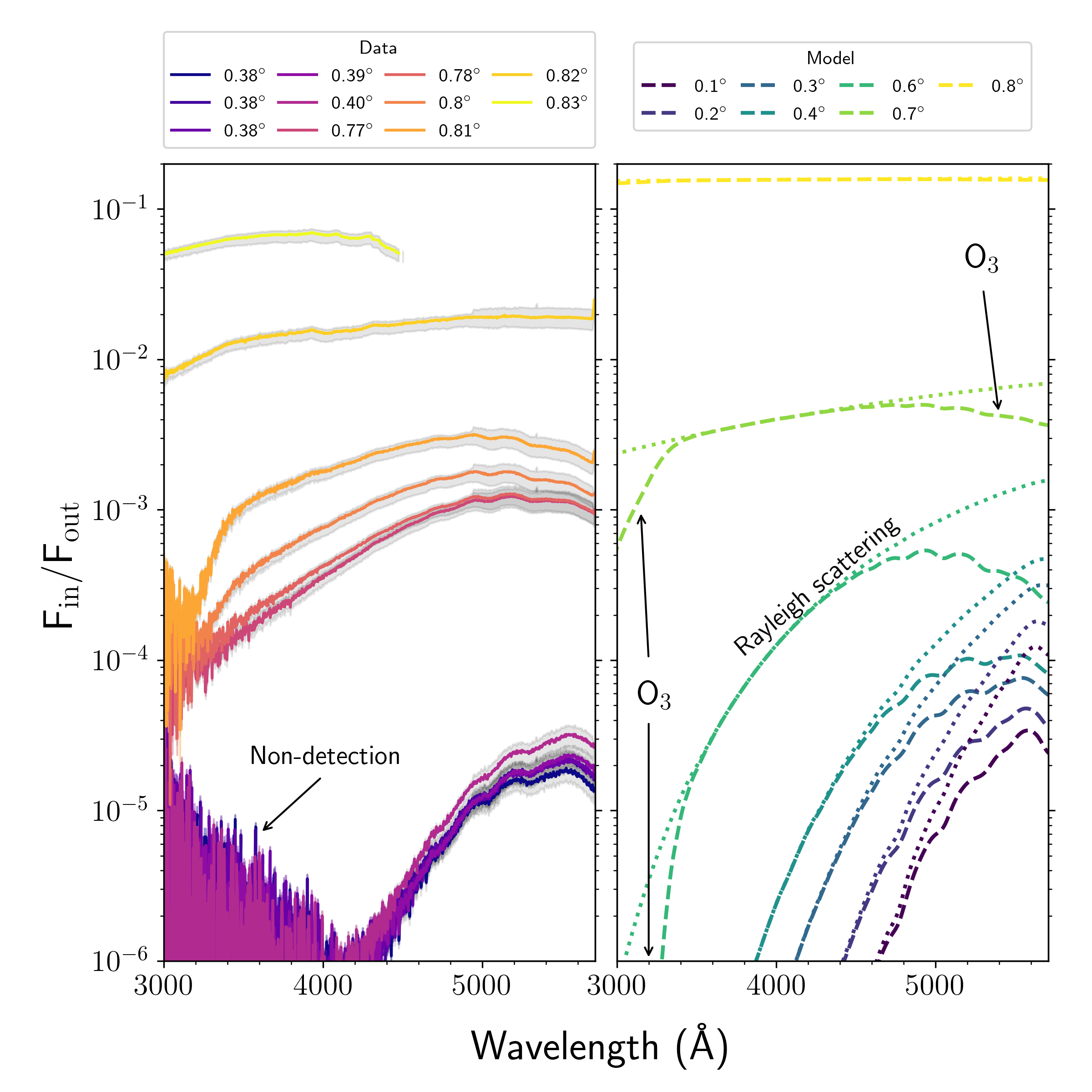}
    \caption{We have combined the transmission spectra of Figure~\ref{fig:G430L_penumbral_transmission} and \ref{fig:G430L_umbral_transmission} into one plot (left panel) and compared to the model transmission spectra from \cite{GarciaMunoz2011} (right panel). The dashed lines represent the model that includes O$_3$, O$_2$, and Rayleigh scattering, and the dotted lines represent the model that excludes O$_3$. Spectral features have been labeled.}
    \label{fig:G430L_penumbral_transpec_withmodel}
\end{figure}

\begin{figure}
    \centering
    \includegraphics[width=0.9\textwidth]{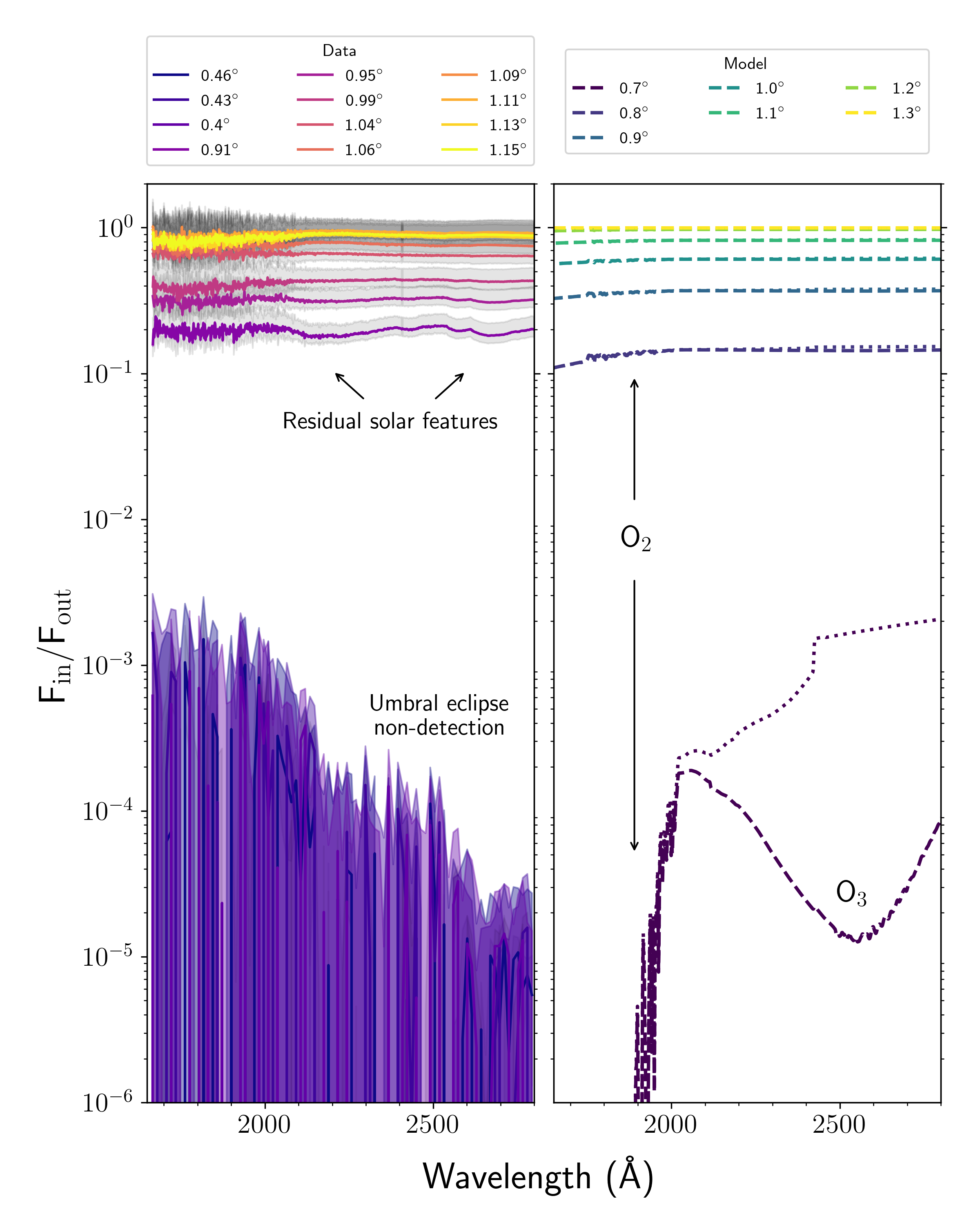}
    \caption{We have combined the transmission spectra of Figure~\ref{fig:G230LB_penumbral_transmission} and the non-detection of the umbral eclipse in the near-UV into one plot (left panel) and compared to the model transmission spectra from \cite{GarciaMunoz2011} (right panel). As in Figure~\ref{fig:G430L_penumbral_transpec_withmodel}, the dashed lines represent the model that includes O$_3$, O$_2$, and Rayleigh scattering, and the dotted lines represent the model that excludes O$_3$, and spectral features have been labeled.  The umbral eclipse spectra have been binned by 10 for clarity.}
    \label{fig:G230LB_penumbral_transpec_withmodel}
\end{figure}

In Figure~\ref{fig:model_data_lightcurve}, we construct broadband transit light curves from our data and the \cite{GarciaMunoz2011} model that show the ratio of in-transit (in-eclipse) flux to out-of-transit (out-of-eclipse) flux as a function of eclipse angle, which is a proxy for time. We note that the minimum eclipse angle of the 21 January 2019 total lunar eclipse was 0.4$^{\circ}$, and future eclipses will vary in their maximum eclipse depth. We find that for any given eclipse angle, the F$_{\rm in}$/F$_{\rm out}$ ratio we observe is less than predicted by the model, meaning that we observe a deeper eclipse. This is consistent with the earlier discussion that the absence of clouds and aerosols in the model (e.g., missing opacity sources make the model F$_{in}$/F$_{out}$ too large), and an imperfect correction for CLV effects, which makes the observed F$_{in}$/F$_{out}$ too small. Following the method of \cite{Keen1983}, we estimate from our optical umbral spectra the average global aerosol loading and compare it to past eclipse observations. Note that we use the optical spectra for this because there are no near-UV eclipse observations to compare to. We compute the theoretical and observed F$_{\rm in}$/F$_{\rm out}$ ratios over a 5400-5600 \AA~bandpass, which serves as a proxy for V band. We then compute the optical depths $\tau=-ln F_{\rm in}/F_{\rm out}$ along Earth's limb for the umbral eclipse observations and the corresponding models, and take the difference ($\tau_{o-c}$), where $o$ is for observations and $c$ is for the calculated model. This eclipse's mean $\tau_{o-c}$ = 1.8$\pm$0.1, however, this value could be too high, because we did not account for CLV effects in our umbral spectra. In order to estimate the magnitude of CLV effects on $\tau_{o-c}$, we consider two possible CLV correction factors: $I/I_0$ = 0.2 and 0.8. For $I/I_0$=0.2 (possibly an extreme value), then $\tau_{o-c}$ becomes 0.2$\pm$0.1, and for $I/I_0$=0.8 (a value applicable to our penumbral spectra), then $\tau_{o-c}$ becomes 1.6$\pm$0.1. However, \cite{Keen1983} did not account for CLV effects in their analysis, so we use $\tau_{o-c}$=1.8$\pm$0.1 as this eclipse's global aerosol loading parameter. Our measured value is slightly elevated above the mean background value 0.7 determined by \cite{Keen1983}, but is significantly lower than the values observed during eclipses that occurred soon after major volcanic eruptions (e.g., $\tau_{o-c}$ = 2-6), and is consistent with the lack of recent major volcanic eruptions on Earth. See \cite{GarciaMunoz2011b} and \cite{Ugolnikov2013} for recent analyses of the impact of volcanic eruptions on the appearance of lunar eclipses.

These eclipse spectra provide a useful ground-truth approximation of what Earth's UV and visible O$_3$ features may look like on transiting exoplanets, but a real Earth analog orbiting a different star would likely appear different.  Transit geometries for exoplanets are affected by the planet-star distance, which is in turn affected by the host star type if the planet must be in the habitable zone, the range of star-planet separations where liquid water can stably exist on the planet's surface \cite{Kasting1993}. Compared to brighter stars, M dwarf stellar hosts offer the best prospects for detecting transiting exoplanets because of their larger transit depths and, for habitable zone planets, more frequent transit events. For planets orbiting M dwarfs, transit geometry and refraction effects mean altitudes down to about 10 km could be probed at continuum wavelengths in the near-IR \citep{Misra2014, Betremieux2014}, but for visible and UV wavelengths, Rayleigh scattering and strong ozone absorption features push the transit baseline to higher altitudes \citep[20-60 km, e.g.,][]{Meadows2018}. 

Photochemistry further complicates the spectral features we could expect to see on exoplanets. The amount of ozone in a planet's atmosphere results from the balance between its production and destruction rates. \citet{Segura2005} considered how ozone is affected by M dwarf photochemistry and found that a modern Earth analog orbiting GJ 643 (M3.5V) might accumulate 50\% more ozone than Earth around the Sun due to the shape of its star's UV spectrum. Recall that O$_3$ forms via O$_2$ photolysis, and GJ 643 emits more photons shortward of 2000 \AA~where O$_2$ is photolyzed compared to the Sun \citep{Segura2005}. Ozone itself is photolyzed by photons at wavelengths between 2000-3100 \AA, where M dwarfs produce fewer photons than the Sun. These photochemical feedbacks can produce complex spectral consequences, affecting the atmospheric abundances of not only ozone, but also other features.  For instance, diminished O$_3$ photolysis around M dwarfs means that significantly more CH$_4$, another biosignature, can accumulate in oxygenated atmospheres, because oxygen radicals produced by O$_3$ photolysis are the dominant sink of CH$_4$ in an Earth-like atmosphere \citep{Segura2005, Meadows2018}.  Thus, a complex web of factors, including photochemistry, orbital properties, and transit geometry, must be considered when anticipating spectral features that might be present on real exoplanets. 

Figure \ref{fig:Earth_transit} shows a comparison of photochemically-self consistent spectra of a modern Earth analog from  \citet{Meadows2018} vs. a simulated spectrum for true Earth around the sun. We have also added two of the \cite{GarciaMunoz2011} model spectra converted to effective height\footnote{We note that, due to the linear scaling and its much larger absolute F$_{in}$/F$_{out}$ values, the $e$=1.0$^{\circ}$ model spectrum appears to have larger amplitude spectral features than the $e$=0.7$^{\circ}$ model spectrum in Figure~\ref{fig:Earth_transit}. When we normalize the model spectra to their mean values, the $e$=0.7$^{\circ}$ model's O$_3$ features are many times larger in amplitude than the $e$=1.0$^{\circ}$ model.} via the method described in Section~\ref{sec:obs} based on Equation 10 of \cite{Vidal-Madjar2010}. We assumed that $e$ is approximately equal to the apparent angular separation of the Earth and Sun and using the Earth-Sun and Earth-moon distances used in that work. We also normalized the model spectra to be 23 km between 4520-4540 \AA~\citep{Arnold2014} as was done for our observed spectra. It is clear that differences in transit geometry and photochemistry drive differing transit heights. Nevertheless, studying Earth as an exoplanet is useful, for it allows spectral models to be tested against real data. In Figure~\ref{fig:transspec_calibrated_km}, we show our calibrated transmission spectra for comparison with the transiting exo-Earth and lunar eclipse models from Figure~\ref{fig:Earth_transit}. There are stark differences between our data and the models at short wavelengths, with our data only reaching effective heights of 25-30 km, rather than 30-40 km. However, the general shapes between 4000-5700 \AA~are in agreement with the models, but especially with the model showing the Earth orbiting the M dwarf Proxima Centauri. It is difficult to rigorously assess the significance or cause of the effective height differences, because of the large uncertainties in our determination of each of relevant parameters in converting to effective height as well as the average eclipse angle and CLV correction corresponding to our data. However, these figures highlight that the rapidly evolving geometry during a transit results in probing different altitudes as a function of time as has been discussed in \cite{GarciaMunoz2011}, \cite{Arnold2014}, and \cite{Kawauchi2018}.


\section{Conclusions} \label{sec:conclusions}

We have presented the first HST observations of a lunar eclipse, and the first glimpse in the near-UV of the Earth as a transiting exoplanet. Observing in the UV requires a space observatory, which is also advantageous for bypassing the path length through Earth's atmosphere between the moon and the Earth. Ground-based observatories must correct for this contamination to isolate the path length along the terminator through Earth's atmosphere between the Sun and the moon. However, observing the moon with HST has different challenges, namely lower spectral resolution and pointing instability. The moon is not a homogeneous surface, and pointing instability does not guarantee that the overall albedo and reflectivity spectrum of the lunar surfaces observed in-transit and out-of-transit are identical. Creating transmission spectra with only Earth's spectral signatures relies on the in-transit and out-of-transit solar and lunar features being identical.  We find evidence for overall albedo variations of the moon between our out-of-eclipse spectra, but defer a thorough analysis to a future paper.

We have constructed normalized irradiance spectra, or transmission spectra, of the Earth for the umbral and penumbral phases of the eclipse. We detect O$_3$ in both the Hartley-Huggins and Chappuis bands during the penumbral phase around 3000-3300 \AA~and 4500-5700 \AA, respectively. During the umbral phase, we detect only the Chappuis band. The other feature that we detect is the Rayleigh scattering slope from 3000-5000 \AA~during both phases of the eclipse. No photons below $\lesssim$4200 \AA~were detected during the umbral phase, which is roughly in agreement with the lunar eclipse models of \cite{GarciaMunoz2011} that show that essentially no light shortward of $\sim$4000 \AA~is transmitted to the lunar surface during a total umbral eclipse. We do not detect any O$_3$ spectral features from 1700-2800 \AA~during the penumbral phase. This is likely because the near-UV observations occurred during late phases of the eclipse (large eclipse angles) when the altitudes probed by refracted sunlight that reach the moon lie above most of the O$_3$ layer and the near-UV spectra have residual solar features larger than the expected amplitude of any O$_3$ features.

Based on the low amplitude of near-UV features in the model spectra at large eclipse angles and the lack of transmitted light at low eclipse angles, the window where near-UV lunar eclipse spectra are useful for diagnosing the Earth's atmosphere is narrow. O$_2$ and O$_3$ absorption is so strong, and the lensing effect which would amplify the signal at small eclipse angles is effectively attenuated by this strong absorption. Also, at these intermediate eclipse angles where O$_3$ signatures might be distinguishable, the normalized irradiance values may approach 10$^{-6}$, the level where the forward-scattered (``diffuse") component of sunlight dominates over the refracted (``direct") component. The diffuse component is roughly wavelength-independent and essentially only contains information about aerosols and gases at very high altitudes, and could mask spectral features of gaseous species at lower altitudes. It would be worthwhile to explore the diffuse component, what it can tell us about aerosols in Earth's and exoplanetary atmospheres, and when it begins to dominate over the direct or refracted sunlight component. The relative timing of our near-UV and optical observations caused us to miss the narrow window of intermediate eclipse angles where the amplitude of near-UV spectral signatures would have been measurable with HST, and a repeat of this experiment could allow us to explore this interesting spectral region further.

Our presented eclipse spectra provide an important ground-truth approximation of what Earth's near-UV and visible O$_3$ may look like on transiting exoplanets. First, photochemistry driven by different stellar radiation environments can strongly impact the chemical makeup of a planet's atmosphere, affecting the relative strengths of spectral features of gases like ozone. Second, the geometry and refraction effects inherent in a lunar eclipse are not the same as a transit observation of a real exoplanet. Many works (e.g., \citealt{Seager2000,Sidis2010,GarciaMunoz2012,Misra2014,Betremieux2016}) have discussed the important effects of refraction and the diversity of transit geometries on the appearance of exoplanet transmission spectroscopy. However, real data of an extreme  geometry like a lunar eclipse is useful for testing models that incorporate these effects. Earth will always be our be studied planet, and by considering solar system bodies like Earth through the lens of analog exoplanets, we will better be able to predict and enhance the capabilities of future exoplanet observing missions, and the models used to study them.

\begin{figure}
    \centering
    \includegraphics[width=\textwidth]{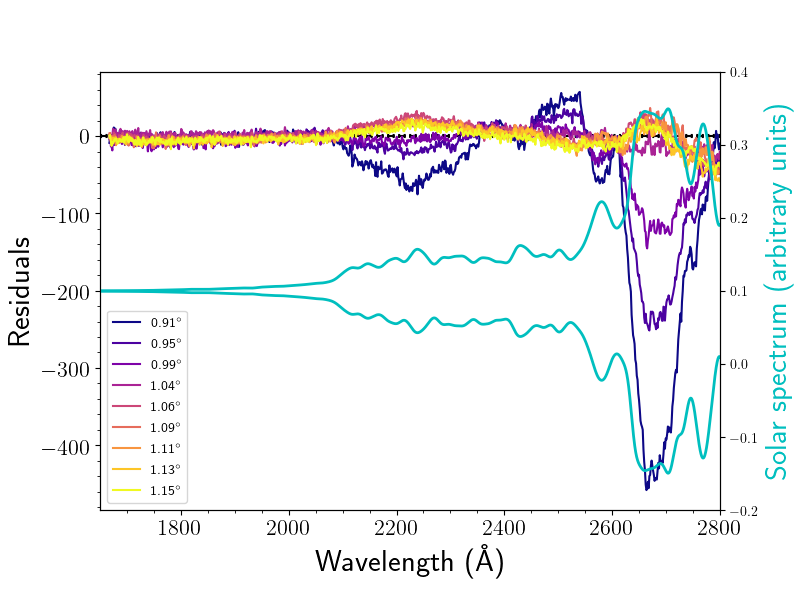}
    \caption{The near-UV transmission spectra are shown as residuals from a straight line, in order to determine their consistency with a flat spectrum. The spectra from Figure~\ref{fig:G230LB_penumbral_transmission} were normalized to unity at 2610 \AA, subtracted from a normalized flat line (equal to unity) and divided by the normalized error bars. Corresponding to the right vertical axis are the solar spectrum and unity minus the solar spectrum, both convolved to match the STIS resolution and plotted in cyan. The spectral features, mostly \ion{Fe}{2} and higher ionization states of Fe, seen in the residuals are consistent with a solar origin.}
    \label{fig:G230LB_residuals_straight_lines}
\end{figure}

\begin{figure}
    \centering
    \includegraphics[width=0.75\textwidth]{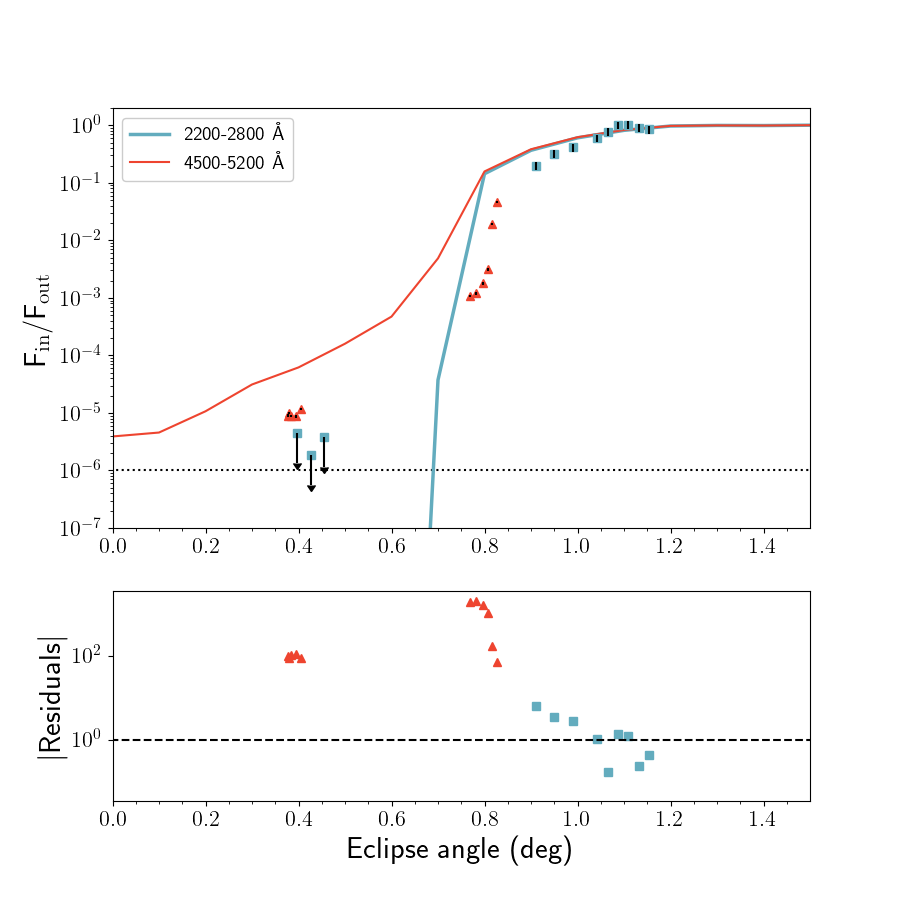}
    \caption{\textit{Top:} Weighted averages of the observed normalized solar flux from 2200-2800 \AA~(squares) and 4500-5700 \AA~(red triangles) are plotted as a function of eclipse angles (error bars on the normalized flux are typically smaller than the size of the data points). The 3 near-UV data points at the smallest observed eclipse angles are non-detections and should be interpreted as 1-$\sigma$~upper limits. Also plotted as solid lines are the model average fluxes over the same wavelength ranges (and in the same color scheme) for comparison. The black error bars represent uncertainties due to albedo and are much larger than the statistical uncertainty error bars, but are still generally the size of the data points. The horizontal dotted line represents a conservative estimate of the forward-scattered (``diffuse") component's strength. \textit{Bottom:} The absolute value of the residuals are shown. The $\pm$1$\sigma$~equivalent is shown as the dashed horizontal line. Points above the line are greater than $\pm$1-$\sigma$ away from the model curve.}
    \label{fig:model_data_lightcurve}
\end{figure}

\begin{figure}
    \centering
    \includegraphics[width=\textwidth]{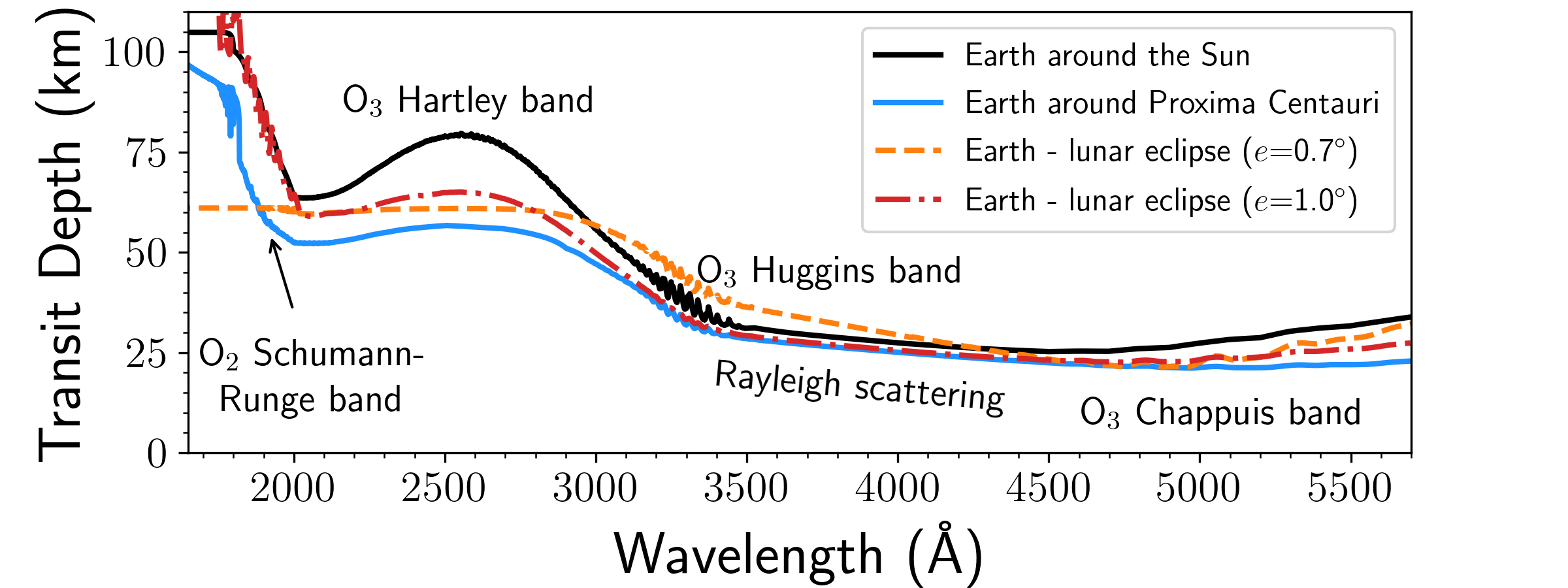}
    \caption{The spectrum of modern Earth as a transiting exoplanet with transit geometry, refraction, and photochemical effects included for Earth around the Sun versus Earth around Proxima Centauri from \citep{Meadows2018}. Absorption and scattering features are labeled. We have also included the $e$=0.7$^{\circ}$ and 1.0$^{\circ}$ from \cite{GarciaMunoz2011} converted into approximate transit depth of the atmosphere assuming the Earth-Sun and Earth-moon separations listed in the text and that the eclipse angle approximately equals the apparent angular separation between the centers of the Earth and Sun. The effective height has been fixed to a mean of 23 km between 4520-4540 \AA~as was done in \cite{Arnold2014}.}
    \label{fig:Earth_transit}
\end{figure}

\begin{figure}
    \centering
    \includegraphics[width=\textwidth]{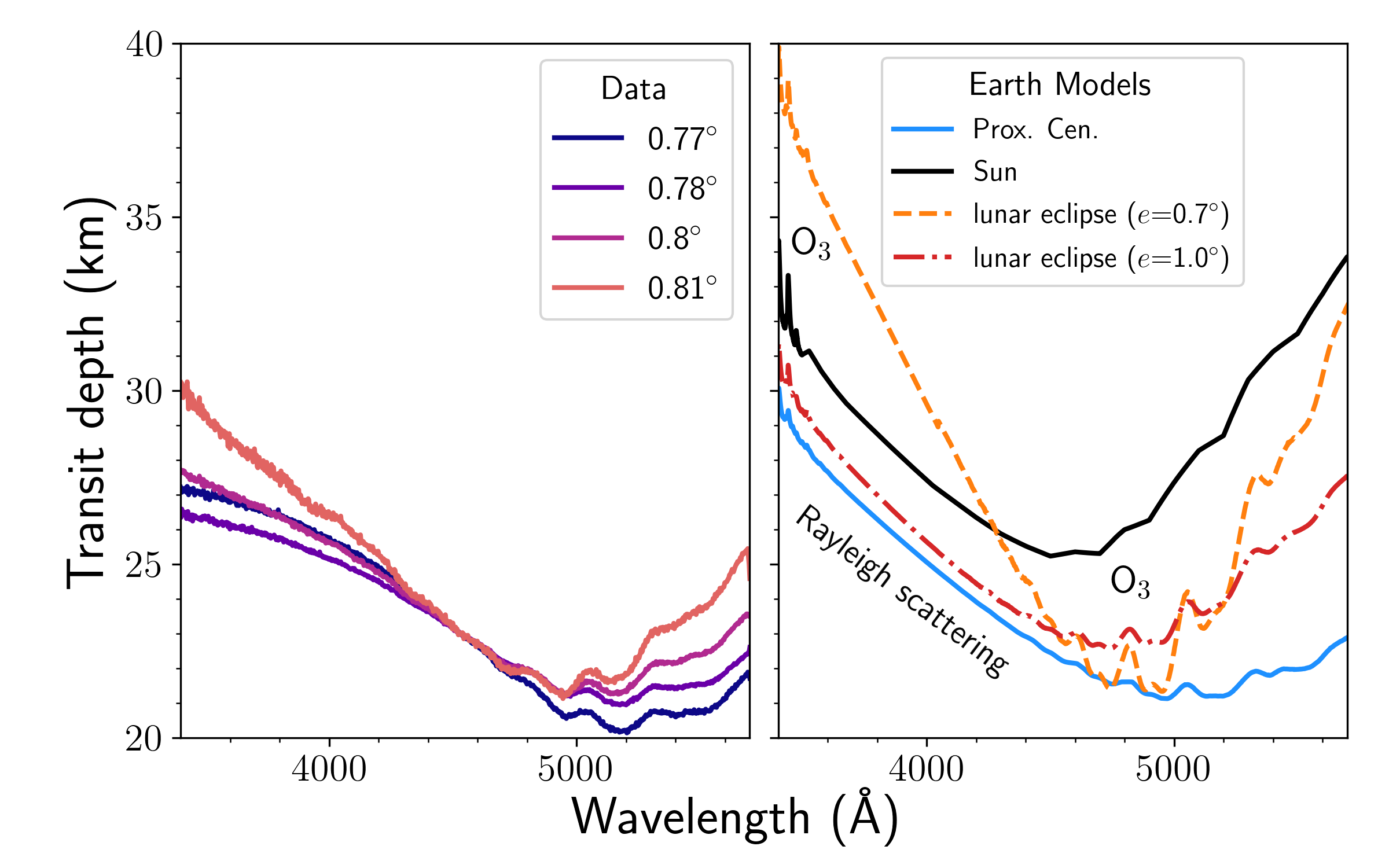}
    \caption{We compare the model transit spectra from Figure~\ref{fig:Earth_transit} \citep{GarciaMunoz2011,Meadows2018} with our roughly calibrated transmission spectra. We have fixed the effective height to be 23 km between 4520-4540 \AA~as was done in \cite{Arnold2014}.}
    \label{fig:transspec_calibrated_km}
\end{figure}

\acknowledgments
We thank the anonymous referee for helpful feedback that improved the quality of this work. The data presented here were obtained as part of the HST Director's Discretionary Time program \#15674. A.Y. acknowledges support by an appointment to the NASA Postdoctoral Program at Goddard Space Flight Center, administered by USRA through a contract with NASA, and STScI grant HST-GO-15674.005. G. A. acknowledges support from the Virtual Planetary Laboratory, supported by the NASA Nexus for Exoplanet System Science (NExSS) research coordination network Grant 80NSSC18K0829, and support from the Goddard Space Flight Center Sellers Exoplanet Environments Collaboration (SEEC), which is funded by the NASA Planetary Science Division's Internal Scientist Funding Model (ISFM). Special thanks to the STScI Director's Office for awarding the observations, and to our program coordinator at STScI, Tony Roman, for doing the hard work of ensuring the success of these observations. We also thank Nat Bachmann and the NAIF team at NASA/JPL for their assistance with SPICE. Thanks to Prabal Saxena, Daniel Moriarty, and Dmitry Vorobiev for helpful discussions regarding lunar reflectance.

\facilities{HST}
\software{Spiceypy\footnote{https://spiceypy.readthedocs.io/en/master/}, cosmics\footnote{https://github.com/ebellm/pyraf-dbsp/blob/master/cosmics.py}, Astropy \citep{astropy2018}, IPython \citep{Perez2007}, Matplotlib \citep{Hunter2007}, Numpy \citep{VanderWalt2011}, Cartopy\footnote{https://scitools.org.uk/cartopy/docs/v0.15/index.html}, Scipy \citep{Millman2011}, stistools\footnote{https://stistools.readthedocs.io/en/latest/} }

\bibliography{ref.bib}{}

\begin{thebibliography}{}
\expandafter\ifx\csname natexlab\endcsname\relax\def\natexlab#1{#1}\fi
\providecommand{\url}[1]{\href{#1}{#1}}
\providecommand{\dodoi}[1]{doi:~\href{http://doi.org/#1}{\nolinkurl{#1}}}
\providecommand{\doeprint}[1]{\href{http://ascl.net/#1}{\nolinkurl{http://ascl.net/#1}}}
\providecommand{\doarXiv}[1]{\href{https://arxiv.org/abs/#1}{\nolinkurl{https://arxiv.org/abs/#1}}}

\bibitem[{Acton {et~al.}(2018)Acton, Bachman, Semenov, \& Wright}]{Acton2018}
Acton, C., Bachman, N., Semenov, B., \& Wright, E. 2018, Planetary and Space
  Science, 150, 9 , \dodoi{https://doi.org/10.1016/j.pss.2017.02.013}

\bibitem[{{Arnold} {et~al.}(2014){Arnold}, {Ehrenreich}, {Vidal-Madjar},
  {Dumusque}, {Nitschelm}, {Querel}, {Hedelt}, {Berthier}, {Lovis}, {Moutou},
  {Ferlet}, \& {Crooker}}]{Arnold2014}
{Arnold}, L., {Ehrenreich}, D., {Vidal-Madjar}, A., {et~al.} 2014, \aap, 564,
  A58, \dodoi{10.1051/0004-6361/201323041}

\bibitem[{{Astropy Collaboration} {et~al.}(2018){Astropy Collaboration},
  {Price-Whelan}, {Sip{\H{o}}cz}, {G{\"u}nther}, {Lim}, {Crawford}, {Conseil},
  {Shupe}, {Craig}, {Dencheva}, {Ginsburg}, {Vand erPlas}, {Bradley},
  {P{\'e}rez-Su{\'a}rez}, {de Val-Borro}, {Aldcroft}, {Cruz}, {Robitaille},
  {Tollerud}, {Ardelean}, {Babej}, {Bach}, {Bachetti}, {Bakanov}, {Bamford},
  {Barentsen}, {Barmby}, {Baumbach}, {Berry}, {Biscani}, {Boquien}, {Bostroem},
  {Bouma}, {Brammer}, {Bray}, {Breytenbach}, {Buddelmeijer}, {Burke},
  {Calderone}, {Cano Rodr{\'\i}guez}, {Cara}, {Cardoso}, {Cheedella}, {Copin},
  {Corrales}, {Crichton}, {D'Avella}, {Deil}, {Depagne}, {Dietrich}, {Donath},
  {Droettboom}, {Earl}, {Erben}, {Fabbro}, {Ferreira}, {Finethy}, {Fox},
  {Garrison}, {Gibbons}, {Goldstein}, {Gommers}, {Greco}, {Greenfield},
  {Groener}, {Grollier}, {Hagen}, {Hirst}, {Homeier}, {Horton}, {Hosseinzadeh},
  {Hu}, {Hunkeler}, {Ivezi{\'c}}, {Jain}, {Jenness}, {Kanarek}, {Kendrew},
  {Kern}, {Kerzendorf}, {Khvalko}, {King}, {Kirkby}, {Kulkarni}, {Kumar},
  {Lee}, {Lenz}, {Littlefair}, {Ma}, {Macleod}, {Mastropietro}, {McCully},
  {Montagnac}, {Morris}, {Mueller}, {Mumford}, {Muna}, {Murphy}, {Nelson},
  {Nguyen}, {Ninan}, {N{\"o}the}, {Ogaz}, {Oh}, {Parejko}, {Parley}, {Pascual},
  {Patil}, {Patil}, {Plunkett}, {Prochaska}, {Rastogi}, {Reddy Janga},
  {Sabater}, {Sakurikar}, {Seifert}, {Sherbert}, {Sherwood-Taylor}, {Shih},
  {Sick}, {Silbiger}, {Singanamalla}, {Singer}, {Sladen}, {Sooley},
  {Sornarajah}, {Streicher}, {Teuben}, {Thomas}, {Tremblay}, {Turner},
  {Terr{\'o}n}, {van Kerkwijk}, {de la Vega}, {Watkins}, {Weaver}, {Whitmore},
  {Woillez}, {Zabalza}, \& {Astropy Contributors}}]{astropy2018}
{Astropy Collaboration}, {Price-Whelan}, A.~M., {Sip{\H{o}}cz}, B.~M., {et~al.}
  2018, \aj, 156, 123, \dodoi{10.3847/1538-3881/aabc4f}

\bibitem[{{B{\'e}tr{\'e}mieux}(2016)}]{Betremieux2016}
{B{\'e}tr{\'e}mieux}, Y. 2016, \mnras, 456, 4051, \dodoi{10.1093/mnras/stv2955}

\bibitem[{B{\'e}tr{\'e}mieux \& Kaltenegger(2014)}]{Betremieux2014}
B{\'e}tr{\'e}mieux, Y., \& Kaltenegger, L. 2014, The Astrophysical Journal,
  791, 7

\bibitem[{Cockell(1998)}]{Cockell1998}
Cockell, C.~S. 1998, Journal of theoretical biology, 193, 717

\bibitem[{{Des Marais} {et~al.}(2002){Des Marais}, {Harwit}, {Jucks},
  {Kasting}, {Lin}, {Lunine}, {Schneider}, {Seager}, {Traub}, \&
  {Woolf}}]{DesMarais2002}
{Des Marais}, D.~J., {Harwit}, M.~O., {Jucks}, K.~W., {et~al.} 2002,
  Astrobiology, 2, 153, \dodoi{10.1089/15311070260192246}

\bibitem[{Eckermann {et~al.}(2007)Eckermann, Broutman, Stollberg, Ma,
  McCormack, \& Hogan}]{Eckermann2007}
Eckermann, S.~D., Broutman, D., Stollberg, M.~T., {et~al.} 2007, Journal of
  Geophysical Research: Atmospheres, 112, \dodoi{10.1029/2006JD007880}

\bibitem[{{Ehrenreich} {et~al.}(2006){Ehrenreich}, {Tinetti}, {Lecavelier Des
  Etangs}, {Vidal-Madjar}, \& {Selsis}}]{Ehrenreich2006}
{Ehrenreich}, D., {Tinetti}, G., {Lecavelier Des Etangs}, A., {Vidal-Madjar},
  A., \& {Selsis}, F. 2006, \aap, 448, 379, \dodoi{10.1051/0004-6361:20053861}

\bibitem[{{Garc{\'\i}a Mu{\~n}oz} \& {Pall{\'e}}(2011)}]{GarciaMunoz2011}
{Garc{\'\i}a Mu{\~n}oz}, A., \& {Pall{\'e}}, E. 2011, \jqsrt, 112, 1609,
  \dodoi{10.1016/j.jqsrt.2011.03.017}

\bibitem[{{Garc{\'\i}a Mu{\~n}oz} {et~al.}(2011){Garc{\'\i}a Mu{\~n}oz},
  {Pall{\'e}}, {Zapatero Osorio}, \& {Mart{\'\i}n}}]{GarciaMunoz2011b}
{Garc{\'\i}a Mu{\~n}oz}, A., {Pall{\'e}}, E., {Zapatero Osorio}, M.~R., \&
  {Mart{\'\i}n}, E.~L. 2011, \grl, 38, L14805, \dodoi{10.1029/2011GL047981}

\bibitem[{{Garc{\'\i}a Mu{\~n}oz} {et~al.}(2012){Garc{\'\i}a Mu{\~n}oz},
  {Zapatero Osorio}, {Barrena}, {Monta{\~n}{\'e}s-Rodr{\'\i}guez},
  {Mart{\'\i}n}, \& {Pall{\'e}}}]{GarciaMunoz2012}
{Garc{\'\i}a Mu{\~n}oz}, A., {Zapatero Osorio}, M.~R., {Barrena}, R., {et~al.}
  2012, \apj, 755, 103, \dodoi{10.1088/0004-637X/755/2/103}

\bibitem[{González-Merino {et~al.}(2013)González-Merino, Pallé, Motalebi,
  Montañés-Rodríguez, \& Kissler-Patig}]{GonzalezMerino2013}
González-Merino, B., Pallé, E., Motalebi, F., Montañés-Rodríguez, P., \&
  Kissler-Patig, M. 2013, Monthly Notices of the Royal Astronomical Society,
  435, 2574, \dodoi{10.1093/mnras/stt1463}

\bibitem[{{Greve} \& {Neckel}(1996)}]{Greve1996}
{Greve}, A., \& {Neckel}, H. 1996, \aaps, 120, 35

\bibitem[{{Hapke}(1981)}]{Hapke1981}
{Hapke}, B. 1981, \jgr, 86, 3039, \dodoi{10.1029/JB086iB04p03039}

\bibitem[{{Hapke} {et~al.}(2012){Hapke}, {Denevi}, {Sato}, {Braden}, \&
  {Robinson}}]{Hapke2012}
{Hapke}, B., {Denevi}, B., {Sato}, H., {Braden}, S., \& {Robinson}, M. 2012,
  Journal of Geophysical Research (Planets), 117, E00H15,
  \dodoi{10.1029/2011JE003916}

\bibitem[{{Hapke}(1963)}]{Hapke1963}
{Hapke}, B.~W. 1963, \jgr, 68, 4571, \dodoi{10.1029/JZ068i015p04571}

\bibitem[{Hegde {et~al.}(2015)Hegde, Paulino-Lima, Kent, Kaltenegger, \&
  Rothschild}]{Hegde2015}
Hegde, S., Paulino-Lima, I.~G., Kent, R., Kaltenegger, L., \& Rothschild, L.
  2015, Proceedings of the National Academy of Sciences, 112, 3886

\bibitem[{{Hestroffer} \& {Magnan}(1998)}]{Hestroffer1998}
{Hestroffer}, D., \& {Magnan}, C. 1998, \aap, 333, 338

\bibitem[{{Hunter}(2007)}]{Hunter2007}
{Hunter}, J.~D. 2007, Computing in Science and Engineering, 9, 90,
  \dodoi{10.1109/MCSE.2007.55}

\bibitem[{{Kasting} {et~al.}(1993){Kasting}, {Whitmire}, \&
  {Reynolds}}]{Kasting1993}
{Kasting}, J.~F., {Whitmire}, D.~P., \& {Reynolds}, R.~T. 1993, \icarus, 101,
  108, \dodoi{10.1006/icar.1993.1010}

\bibitem[{{Kawauchi} {et~al.}(2018){Kawauchi}, {Narita}, {Sato}, {Hirano},
  {Kawashima}, {Nakamoto}, {Yamashita}, \& {Tamura}}]{Kawauchi2018}
{Kawauchi}, K., {Narita}, N., {Sato}, B., {et~al.} 2018, \pasj, 70, 84,
  \dodoi{10.1093/pasj/psy079}

\bibitem[{{Keen}(1983)}]{Keen1983}
{Keen}, R.~A. 1983, Science, 222, 1011, \dodoi{10.1126/science.222.4627.1011}

\bibitem[{{Kiang} {et~al.}(2007{\natexlab{a}}){Kiang}, {Siefert}, {Govindjee},
  \& {Blankenship}}]{Kiang2007a}
{Kiang}, N.~Y., {Siefert}, J., {Govindjee}, \& {Blankenship}, R.~E.
  2007{\natexlab{a}}, Astrobiology, 7, 222, \dodoi{10.1089/ast.2006.0105}

\bibitem[{{Kiang} {et~al.}(2007{\natexlab{b}}){Kiang}, {Segura}, {Tinetti},
  {Govindjee}, {Blankenship}, {Cohen}, {Siefert}, {Crisp}, \&
  {Meadows}}]{Kiang2007b}
{Kiang}, N.~Y., {Segura}, A., {Tinetti}, G., {et~al.} 2007{\natexlab{b}},
  Astrobiology, 7, 252, \dodoi{10.1089/ast.2006.0108}

\bibitem[{{Krissansen-Totton} {et~al.}(2018){Krissansen-Totton}, {Arney}, \&
  {Catling}}]{KrissansenTotton2018}
{Krissansen-Totton}, J., {Arney}, G.~N., \& {Catling}, D.~C. 2018, Proceedings
  of the National Academy of Science, 115, 4105,
  \dodoi{10.1073/pnas.1721296115}

\bibitem[{{McClintock} {et~al.}(2005){McClintock}, {Rottman}, \&
  {Woods}}]{McClintock2005}
{McClintock}, W.~E., {Rottman}, G.~J., \& {Woods}, T.~N. 2005, \solphys, 230,
  225, \dodoi{10.1007/s11207-005-7432-x}

\bibitem[{{Meadows}(2017)}]{Meadows2017}
{Meadows}, V.~S. 2017, Astrobiology, 17, 1022, \dodoi{10.1089/ast.2016.1578}

\bibitem[{Meadows {et~al.}(2018)Meadows, Arney, Schwieterman, Lustig-Yaeger,
  Lincowski, Robinson, Domagal-Goldman, Deitrick, Barnes, Fleming,
  {et~al.}}]{Meadows2018}
Meadows, V.~S., Arney, G.~N., Schwieterman, E.~W., {et~al.} 2018, Astrobiology,
  18, 133

\bibitem[{{Meftah} {et~al.}(2018){Meftah}, {Dam{\'e}}, {Bols{\'e}e},
  {Hauchecorne}, {Pereira}, {Sluse}, {Cessateur}, {Irbah}, {Bureau}, {Weber},
  {Bramstedt}, {Hilbig}, {Thi{\'e}blemont}, {Marchand }, {Lef{\`e}vre},
  {Sarkissian}, \& {Bekki}}]{Meftah2018}
{Meftah}, M., {Dam{\'e}}, L., {Bols{\'e}e}, D., {et~al.} 2018, \aap, 611, A1,
  \dodoi{10.1051/0004-6361/201731316}

\bibitem[{{Millman} \& {Aivazis}(2011)}]{Millman2011}
{Millman}, K.~J., \& {Aivazis}, M. 2011, Computing in Science Engineering, 13,
  9, \dodoi{10.1109/MCSE.2011.36}

\bibitem[{Misra {et~al.}(2014)Misra, Meadows, \& Crisp}]{Misra2014}
Misra, A., Meadows, V., \& Crisp, D. 2014, The Astrophysical Journal, 792, 61

\bibitem[{Montanes-Rodriguez {et~al.}(2006)Montanes-Rodriguez, Pall{\'e},
  Goode, \& Mart{\'\i}n-Torres}]{Montanes2006}
Montanes-Rodriguez, P., Pall{\'e}, E., Goode, P., \& Mart{\'\i}n-Torres, F.
  2006, The Astrophysical Journal, 651, 544

\bibitem[{{Olson} {et~al.}(2018){Olson}, {Schwieterman}, {Reinhard},
  {Ridgwell}, {Kane}, {Meadows}, \& {Lyons}}]{Olson2018}
{Olson}, S.~L., {Schwieterman}, E.~W., {Reinhard}, C.~T., {et~al.} 2018, \apjl,
  858, L14, \dodoi{10.3847/2041-8213/aac171}

\bibitem[{{Pall{\'e}}(2010)}]{Palle2010}
{Pall{\'e}}, E. 2010, in EAS Publications Series, Vol.~41, EAS Publications
  Series, ed. T.~{Montmerle}, D.~{Ehrenreich}, \& A.~M. {Lagrange}, 505--516

\bibitem[{{Pall{\'e}} {et~al.}(2009){Pall{\'e}}, {Zapatero Osorio}, {Barrena},
  {Monta{\~n}{\'e}s-Rodr{\'\i}guez}, \& {Mart{\'\i}n}}]{Palle2009}
{Pall{\'e}}, E., {Zapatero Osorio}, M.~R., {Barrena}, R.,
  {Monta{\~n}{\'e}s-Rodr{\'\i}guez}, P., \& {Mart{\'\i}n}, E.~L. 2009, \nat,
  459, 814, \dodoi{10.1038/nature08050}

\bibitem[{{Perez} \& {Granger}(2007)}]{Perez2007}
{Perez}, F., \& {Granger}, B.~E. 2007, Computing in Science and Engineering, 9,
  21, \dodoi{10.1109/MCSE.2007.53}

\bibitem[{{Pettit} \& {Nicholson}(1930)}]{Pettit1930}
{Pettit}, E., \& {Nicholson}, S.~B. 1930, \apj, 71, 102, \dodoi{10.1086/143236}

\bibitem[{{Planavsky} {et~al.}(2014){Planavsky}, {Reinhard}, {Wang}, {Thomson},
  {McGoldrick}, {Rainbird}, {Johnson}, {Fischer}, \& {Lyons}}]{Planavsky2014}
{Planavsky}, N.~J., {Reinhard}, C.~T., {Wang}, X., {et~al.} 2014, Science, 346,
  635, \dodoi{10.1126/science.1258410}

\bibitem[{Reinhard {et~al.}(2017)Reinhard, Olson, Schwieterman, \&
  Lyons}]{Reinhard2017}
Reinhard, C.~T., Olson, S.~L., Schwieterman, E.~W., \& Lyons, T.~W. 2017,
  Astrobiology, 17, 287

\bibitem[{{Robinson} {et~al.}(2011){Robinson}, {Meadows}, {Crisp}, {Deming},
  {A'Hearn}, {Charbonneau}, {Livengood}, {Seager}, {Barry}, {Hearty},
  {Hewagama}, {Lisse}, {McFadden}, \& {Wellnitz}}]{Robinson2011}
{Robinson}, T.~D., {Meadows}, V.~S., {Crisp}, D., {et~al.} 2011, Astrobiology,
  11, 393, \dodoi{10.1089/ast.2011.0642}

\bibitem[{{Sagan} {et~al.}(1993){Sagan}, {Thompson}, {Carlson}, {Gurnett}, \&
  {Hord}}]{Sagan1993}
{Sagan}, C., {Thompson}, W.~R., {Carlson}, R., {Gurnett}, D., \& {Hord}, C.
  1993, \nat, 365, 715, \dodoi{10.1038/365715a0}

\bibitem[{{Schwieterman} {et~al.}(2015){Schwieterman}, {Cockell}, \&
  {Meadows}}]{Schwieterman2015}
{Schwieterman}, E.~W., {Cockell}, C.~S., \& {Meadows}, V.~S. 2015,
  Astrobiology, 15, 341, \dodoi{10.1089/ast.2014.1178}

\bibitem[{{Schwieterman} {et~al.}(2018){Schwieterman}, {Kiang}, {Parenteau},
  {Harman}, {DasSarma}, {Fisher}, {Arney}, {Hartnett}, {Reinhard}, {Olson},
  {Meadows}, {Cockell}, {Walker}, {Grenfell}, {Hegde}, {Rugheimer}, {Hu}, \&
  {Lyons}}]{Schwieterman2018}
{Schwieterman}, E.~W., {Kiang}, N.~Y., {Parenteau}, M.~N., {et~al.} 2018,
  Astrobiology, 18, 663, \dodoi{10.1089/ast.2017.1729}

\bibitem[{{Seager} \& {Sasselov}(2000)}]{Seager2000}
{Seager}, S., \& {Sasselov}, D.~D. 2000, \apj, 537, 916, \dodoi{10.1086/309088}

\bibitem[{Segura {et~al.}(2005)Segura, Kasting, Meadows, Cohen, Scalo, Crisp,
  Butler, \& Tinetti}]{Segura2005}
Segura, A., Kasting, J.~F., Meadows, V., {et~al.} 2005, Astrobiology, 5, 706

\bibitem[{{Sidis} \& {Sari}(2010)}]{Sidis2010}
{Sidis}, O., \& {Sari}, R. 2010, \apj, 720, 904,
  \dodoi{10.1088/0004-637X/720/1/904}

\bibitem[{{Strassmeier} {et~al.}(2020){Strassmeier}, {Ilyin}, {Keles},
  {Mallonn}, {J{\"a}rvinen}, {Weber}, {Mackebrandt}, \&
  {Hill}}]{Strassmeier2020}
{Strassmeier}, K.~G., {Ilyin}, I., {Keles}, E., {et~al.} 2020, \aap, 635, A156,
  \dodoi{10.1051/0004-6361/201936091}

\bibitem[{{Ugolnikov} {et~al.}(2013){Ugolnikov}, {Punanova}, \&
  {Krushinsky}}]{Ugolnikov2013}
{Ugolnikov}, O.~S., {Punanova}, A.~F., \& {Krushinsky}, V.~V. 2013, \jqsrt,
  116, 67, \dodoi{10.1016/j.jqsrt.2012.11.013}

\bibitem[{{van der Walt} {et~al.}(2011){van der Walt}, {Colbert}, \&
  {Varoquaux}}]{VanderWalt2011}
{van der Walt}, S., {Colbert}, S.~C., \& {Varoquaux}, G. 2011, Computing in
  Science and Engineering, 13, 22, \dodoi{10.1109/MCSE.2011.37}

\bibitem[{{van Dokkum}(2001)}]{vanDokkum2001}
{van Dokkum}, P.~G. 2001, Publications of the Astronomical Society of the
  Pacific, 113, 1420, \dodoi{10.1086/323894}

\bibitem[{{Vidal-Madjar} {et~al.}(2010){Vidal-Madjar}, {Arnold}, {Ehrenreich},
  {Ferlet}, {Lecavelier Des Etangs}, {Bouchy}, {Segransan}, {Boisse},
  {H{\'e}brard}, {Moutou}, {D{\'e}sert}, {Sing}, {Cabanac}, {Nitschelm},
  {Bonfils}, {Delfosse}, {Desort}, {Diaz}, {Eggenberger}, {Forveille},
  {Lagrange}, {Lovis}, {Pepe}, {Perrier}, {Pont}, {Santos}, \&
  {Udry}}]{Vidal-Madjar2010}
{Vidal-Madjar}, A., {Arnold}, L., {Ehrenreich}, D., {et~al.} 2010, \aap, 523,
  A57, \dodoi{10.1051/0004-6361/201014751}

\bibitem[{{Yan} {et~al.}(2015{\natexlab{a}}){Yan}, {Fosbury}, {Petr-Gotzens},
  {Zhao}, \& {Pall{\'e}}}]{Yan2015a}
{Yan}, F., {Fosbury}, R.~A.~E., {Petr-Gotzens}, M.~G., {Zhao}, G., \&
  {Pall{\'e}}, E. 2015{\natexlab{a}}, \aap, 574, A94,
  \dodoi{10.1051/0004-6361/201425220}

\bibitem[{{Yan} {et~al.}(2015{\natexlab{b}}){Yan}, {Fosbury}, {Petr-Gotzens},
  {Zhao}, \& {Pall{\'e}}}]{Yan2015_CLV}
---. 2015{\natexlab{b}}, \aap, 574, A94, \dodoi{10.1051/0004-6361/201425220}

\bibitem[{{Yan} {et~al.}(2015{\natexlab{c}}){Yan}, {Fosbury}, {Petr-Gotzens},
  {Zhao}, {Wang}, {Wang}, {Liu}, \& {Pall{\'e}}}]{Yan2015b}
{Yan}, F., {Fosbury}, R.~A.~E., {Petr-Gotzens}, M.~G., {et~al.}
  2015{\natexlab{c}}, International Journal of Astrobiology, 14, 255,
  \dodoi{10.1017/S1473550414000172}

\end{thebibliography}
\bibliographystyle{aasjournal}

\clearpage
\clearpage

\startlongtable 
\begin{deluxetable}{ccccccccc}
\tablecolumns{9}
\tablewidth{0pt}
\tablecaption{ Observations  \label{table:obs}} 
\tablehead{\colhead{Rootname} &
           \colhead{Exposure name} &
           \colhead{Eclipse phase} &
           \colhead{Mid Time} & 
           \colhead{t$_{exp}$ (s)} &
           \colhead{Grating} & 
           \colhead{e ($^{\circ}$)} & 
           \colhead{lon ($^{\circ}$)} &
           \colhead{lat ($^{\circ}$)}
           }
\startdata
\multicolumn{9}{c}{\textit{Track 1 begins 4:40:37}} \\
ody902010 & ody902ahq & umbral & 4:47:02 & 293.3 & G230LB & 0.46 & 11.1 & -27.4 \\
 & ody902aiq & umbral & 4:52:40 & 293.3 & G230LB & 0.43 & 4.2 & -20.5 \\
\multicolumn{9}{c}{\textit{Track 2 begins 4:57:54}} \\
ody902020 & ody902ajq & umbral & 5:00:26 & 293.3 & G230LB & 0.40 & -6.8 & -8.5 \\
\multicolumn{9}{c}{\textit{Track 3 begins 5:06:13}} \\
ody902030 & ody902alq & umbral & 5:12:20 & 154.7 & G430L & 0.38 & -21.4 & -1.0 \\
 & ody902amq & umbral & 5:15:39 & 154.7 & G430L & 0.38 & -21.3 & -0.8 \\
 & ody902aoq & umbral & 5:18:58 & 154.7 & G430L & 0.38 & -14.3 & -4.2 \\
 \multicolumn{9}{c}{\textit{Track 4 begins 5:22:37}} \\
ody902040 & ody902apq & umbral & 5:24:00 & 154.8 & G430L & 0.39 & -9.4 & -4.5 \\
 & ody902aqq & umbral & 5:27:19 & 154.8 & G430L & 0.40 & -10.6 & -4.2 \\
 \hline 
 \multicolumn{9}{c}{\textit{Track 5 begins 06:16:05}} \\
ody9a2010 & ody9a2atq & penumbral & 6:20:28 & 48.3 & G430L & 0.77 & -7.6 & -16.6 \\
 & ody9a2avq & penumbral & 6:22:01 & 48.3 & G430L & 0.78 & 4.2 & -20.1 \\
 & ody9a2axq & penumbral & 6:23:34 & 48.3 & G430L & 0.80 & 4.1 & -19.4 \\
ody9a2020 & ody9a2ayq & penumbral & 6:24:52 & 20 & G430L & 0.81 & 2.1 & -17.3 \\
 & ody9a2azq & penumbral & 6:25:56 & 20 & G430L & 0.82 & -0.3 & -14.7 \\
 & ody9a2b0q & penumbral & 6:27:00 & 20 & G430L & 0.83 & -3.1 & -11.4 \\
 \multicolumn{9}{c}{\textit{Track 6 begins 6:29:44}} \\
ody9a2030 & ody9a2b2q & penumbral & 6:36:26 & 225 & G230LB & 0.91 & -8.7 & -17.7 \\
 & ody9a2b4q & penumbral & 6:40:55 & 225 & G230LB & 0.95 & -10.8 & -16.1 \\
 & ody9a2b5q & penumbral & 6:45:24 & 225 & G230LB & 0.99 & -9.3 & -11.64 \\
  \multicolumn{9}{c}{\textit{Track 7 begins 6:50:06}} \\
ody9a2040 & ody9a2b6q & penumbral & 6:51:01 & 100 & G230LB & 1.04 & -11.6 & -3.1 \\
 & ody9a2b7q & penumbral & 6:53:25 & 100 & G230LB & 1.06 & -20.5 & 3.1 \\
 & ody9a2b9q & penumbral & 6:55:49 & 100 & G230LB & 1.09 & -25.8 & 6.8 \\
 & ody9a2baq & penumbral & 6:58:13 & 100 & G230LB & 1.11 & -26.8 & 7.6 \\
 & ody9a2bbq & penumbral & 7:00:37 & 100 & G230LB & 1.13 & -23.0 & 5.0 \\
 & ody9a2bcq & penumbral & 7:03:01 & 100 & G230LB & 1.15 & -15.0 & -1.2 \\
\hline
 \multicolumn{9}{c}{\textit{Track 8 begins 9:27:03}} \\
ody904010 & ody904blq & full moon & 9:31:35 & 45 & G230LB & 2.59 & 6.0 & -21.1 \\
 & ody904bnq & full moon & 9:33:04 & 45 & G230LB & 2.60 & 7.1 & -22.3 \\
ody904020 & ody904bpq & full moon & 9:34:33 & 45 & G230LB & 2.62 & 6.6 & -21.9 \\
 & ody904brq & full moon & 9:36:02 & 45 & G230LB & 2.63 & 4.5 & -20.0 \\
ody904030 & ody904btq & full moon & 9:37:31 & 45 & G230LB & 2.64 & 1.3 & -16.5 \\
 & ody904bvq & full moon & 9:39:00 & 45 & G230LB & 2.66 & -3.0 & -11.6 \\
  \multicolumn{9}{c}{\textit{Track 9 begins 9:41:45}} \\
ody904040 & ody904bxq & full moon & 9:42:25 & 45 & G230LB & 2.69 & -7.7 & -7.4 \\
 & ody904bzq & full moon & 9:43:54 & 45 & G230LB & 2.71 & -6.4 & -9.8 \\
ody904050 & ody904c1q & full moon  & 9:45:23 & 45 & G230LB & 2.72 & -5.6 & -11.2 \\
 & ody904c3q & full moon & 9:46:52 & 45 & G230LB & 2.74 & -5.2 & -11.7 \\
ody904060 & ody904c5q & full moon & 9:48:23 & 50 & G230LB & 2.75 & -5.1 & -11.4 \\
 & ody904c6q & full moon & 9:49:57 & 50 & G230LB & 2.77 & -4.9 & -10.3 \\
   \multicolumn{9}{c}{\textit{Track 10 begins 9:53:28}} \\
ody904070 & ody904c7q & full moon & 9:54:22 & 75 & G230LB & 2.81 & -10.5 & -4.6 \\
 & ody904c9q & full moon & 9:56:21 & 75 & G230LB & 2.83 & -14.3 & -3.2 \\
ody904080 & ody904ccq & full moon & 10:02:36 & 0.5 & G430L & 2.89 & -15.6 & -3.0 \\
ody904090 & ody904cdq & full moon & 10:03:21 & 0.5 & G430L & 2.90 & -14.3 & -3.7 \\
ody9040A0 & ody904ceq & full moon & 10:04:06 & 0.5 & G430L & 2.91 & -12.7 & -4.6 \\
ody9040B0 & ody904cfq & full moon & 10:04:51 & 0.5 & G430L & 2.91 & -10.7 & -5.8 \\
ody9040C0 & ody904cgq & full moon & 10:05:36 & 0.5 & G430L & 2.92 & -8.4 & -7.3 \\
  \multicolumn{9}{c}{\textit{Track 11 begins 10:09:07}} \\
ody9040D0 & ody904ciq & full moon & 10:09:24 & 0.7 & G430L & 2.96 & -7.3 & -6.1 \\
ody9040E0 & ody904cjq & full moon & 10:10:09 & 0.7 & G430L & 2.96 & -8.5 & -5.3 \\
ody9040F0 & ody904ckq & full moon & 10:10:54 & 0.7 & G430L & 2.97 & -9.3 & -4.8 \\
ody9040G0 & ody904clq & full moon & 10:11:39 & 0.7 & G430L & 2.98 & -9.6 & -4.7 \\
ody9040H0 & ody904cmq & full moon & 10:12:24 & 0.7 & G430L & 2.99 & -9.6 & -4.9 \\
ody9040I0 & ody904cnq & full moon & 10:13:09 & 0.7 & G430L & 2.99 & -9.1 & -5.4\\
ody9040J0 & ody904coq & full moon & 10:13:54 & 0.7 & G430L & 3.00 & -8.3 & -6.4 \\
\enddata
\tablecomments{Times are UTC on 2019-Jan-21. $e$ is the eclipse angle in degrees evaluated for lon=0$^{\circ}$, lat=0$^{\circ}$ at the exposure mid time, and lon and lat represent the lunar longitude and latitude in degrees of the linear tracks at exposure mid time. Linear track times are also shown. The 52\arcsec$\times$2\arcsec~slit was used in all exposures.}
\end{deluxetable}

\listofchanges

\end{document}